\documentclass[aps,pre,twocolumn,superscriptaddress,showpacs]{revtex4-1}
\usepackage{amsmath}
\usepackage{amssymb}
\usepackage{graphicx}% Include figure files
\usepackage{dcolumn}% Align table columns on decimal point
\usepackage{bm}% bold math
\usepackage{psfrag}

\newcommand{\refsec}[1]{\mbox{Sec.~\ref{#1}}}

\begin{document}
\title{Algebraic theory of crystal vibrations: Singularities and zeros in
vibrations of 1D and 2D lattices}
\author{F.~Iachello} \email{francesco.iachello@yale.edu}
\affiliation{Center for Theoretical Physics, Sloane Physics Laboratory, Yale University, New Haven, CT 06520-8120, USA}
\author{B.~Dietz} \email{dietz@ikp.tu-darmstadt.de}
\affiliation{Institut f{\"u}r Kernphysik, Technische Universit{\"a}t
Darmstadt, D-64289 Darmstadt, Germany}
\author{M.~Miski-Oglu}
\affiliation{Institut f{\"u}r Kernphysik, Technische Universit{\"a}t
Darmstadt, D-64289 Darmstadt, Germany}
\author{A.~Richter}
\affiliation{Institut f{\"u}r
Kernphysik, Technische Universit{\"a}t Darmstadt, D-64289 Darmstadt,
Germany}
%\maketitle
\date{\today}% It is always \today, today,
\begin{abstract}
A novel method for the calculation of the energy dispersion relation (EDR)
and density of states (DOS) in one (1D) and two (2D) dimensions is introduced and
applied to linear lattices (1D) and square and hexagonal lattices (2D). The
(van Hove) singularities and (Dirac) zeros of the DOS are discussed. Results
for the 2D hexagonal lattice (graphene-like materials) are compared with
experimental data in microwave photonic crystals.
\end{abstract}


\maketitle

\section{Introduction\label{Intro}}

Phonon energy dispersion relations (EDR) and the density of states (DOS) are a
fundamental physical property of a solid especially for determining the
mechanical, thermal and other condensed matter phenomena. Although most of
the interest in the past has been in three dimensional (3D) systems, for
example diamond, with the discovery of graphene \cite{geim} interest in
systems of lower dimensionality has arisen, including graphene (2D)~\cite{Beenakker2008,castro}, graphene tubules (1D)~\cite{saito} and fullerenes (0D)~\cite{andreoni}.

At the same time, the development of superconducting microwave billiards 
\cite{billiards,billiards2} and of photonic crystals \cite{crystals,crystals2} combined into "microwave photonic crystals", or generally, of "artificial graphene"~\cite{Bittner2010,Kuhl2010,Singha2011,Gomes2012,Tarruell2012,Uehlinger2013,Polini2013a,Bellec2013,Rechtsman2013,Khanikaev2013,Dietz2015}, has opened the way for simulations of crystal vibrations in systems of lower dimensionality (2D, 1D, 0D) with a finite number of units.

Calculations of the EDR and the DOS in infinite systems have been done using a variety
of methods, including tight-binding force-constant models \cite{wallace,Reich2002,saito} and
molecular dynamics methods. Here we introduce a new method, based on the
algebraic theory of molecules \cite{iac-lev}, which reproduces all
previously known results for the fundamental vibration $v=1$ of crystals,
and, in addition, can be extended to overtone vibrations. The method is
particularly well suited for analyzing the EDR and DOS of microwave photonic
crystals.

With this method, we calculate the EDR and DOS of the fundamental ($v=1$)
and overtone ($v=2$) transverse vibrations of 1D and 2D crystals. For $v=1$
we also rederive the EDR and DOS analytically, determining the location of
(van Hove) singularities and (Dirac) zeros in transverse vibrations of 1D
crystals and 2D crystals with square and hexagonal lattices. The latter is
then compared with experimental results in microwave billiards. We are then 
able to determine the scaling behavior of the singularities.

\section{Algebraic method\label{Method}}

\subsection{One-atom\label{1Atom}}

\begin{figure}[h!]
  {\includegraphics[width=0.8\linewidth]{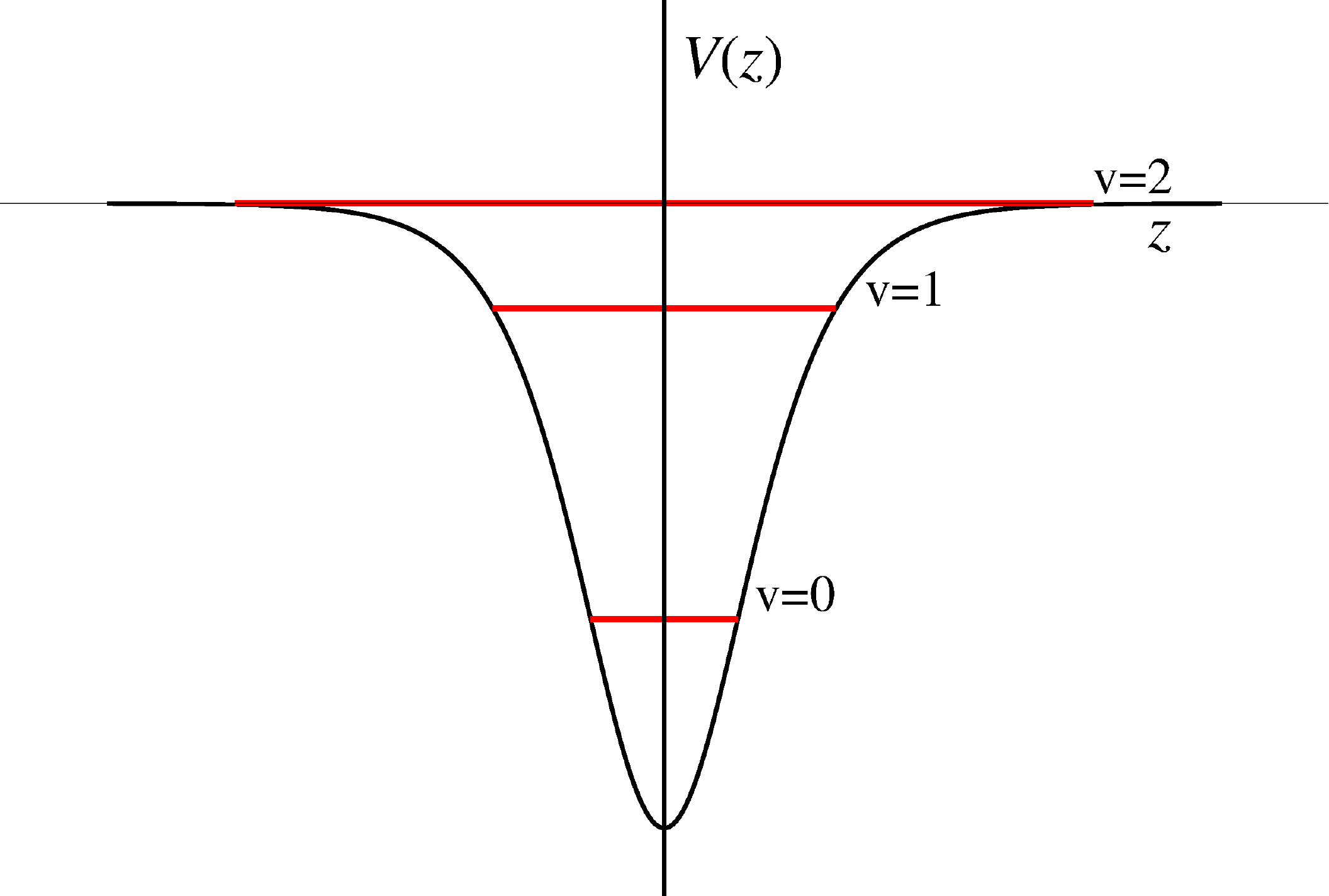}}
  \caption{The P\"{o}schl-Teller potential, Eq.~(\ref{eqn1}), with $\alpha =1,\, V_{0}=1,\, \lambda =3$.}
\label{fig1}
\end{figure}

We consider a one-dimensional (1d) vibration of an atom (for convenience in
the $z$-direction), and assume that it is bound by a potential%
\begin{equation}
V(z)=-\frac{V_{0}\lambda \left( \lambda -1\right) }{\cosh ^{2}\alpha z},
\label{eqn1}
\end{equation}%
as shown in Fig.~\ref{fig1}.
The energy levels in this potential can be described by the simple formula 
\cite{flugge}%
\begin{equation}
E(v)=E_{0}+A\left( v-\xi v^{2}\right) ,
\label{eqn2}
\end{equation}%
where $\xi =\frac{1}{2(\lambda -1)}$ (also written, for integer $\lambda$, as $\frac{1}{N}$) describes the anharmonicity, $A=2V_{0}(\lambda
-1)$ (also written as $\hbar \omega $) the oscillator frequency, and $%
E_{0}=-V_{0}(\lambda -1)^{2}$ the zero-point energy. It was shown in \cite%
{iac-alhassid} that the P\"{o}schl-Teller potential Eq.~(\ref{eqn1}) can be associated
with the algebra of $g\equiv u(2)$. Its eigenstates are described by
representations of $u(2)\supset so(2)$, labeled by $\left\vert
N,v\right\rangle $ where $v=0,1,...,\frac{N}{2}$ or $\frac{N-1}{2}$ (for $N$%
=even or odd) is the vibrational quantum number, and $N$ is the so-called
vibron number which determines how many bound states are in the potential.
The Hamiltonian can then be written in algebraic form as%
\begin{equation}
H=E_{0}+AC,
\label{eqn3}
\end{equation}%
where $E_{0}$ is the zero point energy, $A$ a scale and the operator $C$,
called Casimir operator, has eigenvalues%
\begin{equation}
\left\langle N,v\left\vert C\right\vert N,v\right\rangle =-4\left( v-\frac{%
v^{2}}{N}\right) .
\label{eqn4}
\end{equation}%
From Eq.~(\ref{eqn4}), one can see that $N$ controls the anharmonicity of the
potential $\xi =\frac{1}{N}$. For $N\rightarrow \infty $, the spectrum is
harmonic and has an infinite number of bound states, $v=0,1,...,\infty $.
The potential Eq.~(\ref{eqn1}) is therefore well suited to describe both harmonic and
anharmonic vibrations.

\subsection{Many-atoms\label{MAtoms}}

In the case of many atoms at location $i=1,...,n$, bound by P\"{o}%
schl-Teller potentials, the algebra is $g\equiv u_{1}(2)\oplus
u_{2}(2)\oplus ...\oplus u_{i}(2)\oplus ...\oplus u_{n}(2)$ \cite{iac-oss2,iac-oss1}. The algebraic Hamiltonian for $n$ uncoupled oscillators is 
\begin{equation}
H=E_{0}+\sum_{i=1}^{n}A_{i}C_{i},
\label{eqn5}
\end{equation}%
where $E_{0}$ is the zero-point energy and $A_{i}$ a scale. Its eigenvalues
are%
\begin{equation}
E=E_{0}-4\sum_{i=1}^{n}A_i\left( v_{i}-\frac{v_{i}^{2}}{N_{i}}\right)\, .
\label{eqn6}
\end{equation}
\begin{figure}[h!]
{\includegraphics[width=\linewidth]{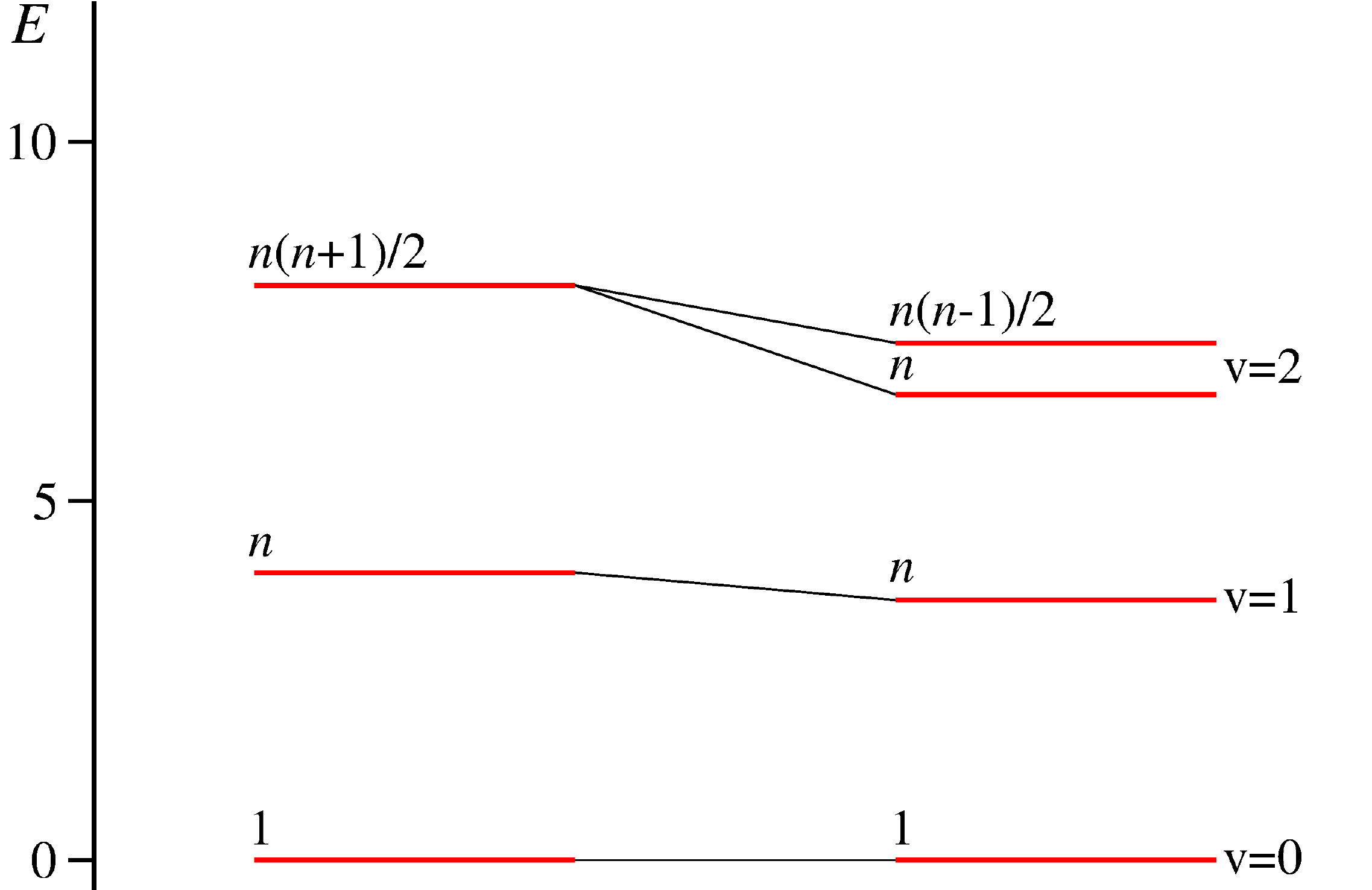}}
\caption{Spectrum of equivalent harmonic oscillators with diagonal
interaction, Eq.~(\ref{eqn11}), with $E_0=0,\, A=-1,\, N=10,\, A^\prime=-0.1$.}
\label{fig2}
\end{figure}
If all the oscillators are equivalent, $A_{i}\equiv A$, $N_{i}=N$, and 
\begin{equation}
E=E_{0}-4A\sum_{i=1}^{n}\left( v_{i}-\frac{v_{i}^{2}}{N}\right) .
\label{eqn7}
\end{equation}%
In the harmonic limit, $N_{i}\rightarrow \infty $, the matrix elements of
the operators $C_{i}$ are the familiar matrix elements of the number
operator $\hat{n}_{i}=b_{i}^{\dag }b_{i}$, where $b_{i}^{\dag }(b_{i})$ are
boson creation and annihilation operators of a phonon at site $i$,
multiplied by $-4$.

The phonons (vibrons) interact with a diagonal and an off-diagonal
interaction. The diagonal interaction is written in terms of the normalized
Casimir operator $C_{ij}$ of the combined $so_{i}(2)\oplus so_{j}(2)$
algebra as%
\begin{eqnarray}
&&\left\langle N_{i},v_{i};N_{j},v_{j}\left\vert C_{ij}\right\vert
N_{i},v_{i};N_{j},v_{j}\right\rangle\\
&=&-4\left[ \left( v_{i}+v_{j}\right) -%
\frac{\left( v_{i}+v_{j}\right) ^{2}}{(N_{i}+N_{j})}\right].\nonumber
\label{eqn8}
\end{eqnarray}%
In practical calculations, it is convenient to subtract from $C_{ij}$ a
contribution that can be absorbed in the Casimir operators of the individual
modes $i$ and $j$ given in Eq.~(\ref{eqn4}), thus considering an operator $%
C_{ij}^{\prime }$ whose matrix elements are%
\begin{eqnarray}
&&\left\langle N_{i},v_{i};N_{j},v_{j}\left\vert C_{ij}^{\prime }\right\vert
N_{i},v_{i};N_{j},v_{j}\right\rangle\\
&=&-4\left[ \frac{v_{i}^{2}}{N_{i}}+\frac{%
v_{j}^{2}}{N_{j}}-\frac{\left( v_{i}+v_{j}\right) ^{2}}{\left(
N_{i}+N_{j}\right) }\right].\nonumber
\label{eqn9}
\end{eqnarray}%
The Hamiltonian for $n$ oscillators diagonally coupled is 
\begin{equation}
H=E_{0}+\sum_{i=1}^{n}A_{i}C_{i}+\sum_{i\neq j}^{n}A_{ij}^{\prime
}C_{ij}^{\prime }.
\label{eqn10}
\end{equation}%
If all oscillators are equivalent, $N_{i}=N$, $A_{i}=A$, $A_{ij}^{\prime
}=A^{\prime }$, and the eigenvalues are%
\begin{eqnarray}
E&=&E_{0}-4A\sum_{i=1}^{n}\left( v_{i}-\frac{v_{i}^{2}}{N}\right)\nonumber\\
&-&4A^{\prime}\sum_{i\neq j}^{n}\left[ \frac{v_{i}^{2}}{N}+\frac{v_{j}^{2}}{N}
-\frac{\left( v_{i}+v_{j}\right) ^{2}}{2N}\right] .
\label{eqn11}
\end{eqnarray}%
The diagonal interaction represents the modification in the P\"{o}%
schl-Teller potentials at each site due to the other sites. It is usually
small and it can be neglected. The spectrum of states of equivalent
oscillators with diagonal interaction is shown in Fig.~\ref{fig2} for up to two phonons, $%
\sum_{i}v_{i}=2$, \ both for harmonic ($N\rightarrow \infty $) and
anharmonic oscillators. The degree of degeneracy of the levels is given below 
in Eq.~(\ref{eqn18}) and is shown to the left of the levels.

The off-diagonal interaction is written in terms of operators, called
Majorana operators $M_{ij}$, with matrix elements~\cite{iac-lev,iac-oss2,iac-oss1} 
\begin{eqnarray}
&&\left\langle N_{i},v_{i}+1;N_{j},v_{j}-1\left\vert M_{ij}\right\vert
N_{i},v_{i};N_{j},v_{j}\right\rangle\\
&=&-\left[ v_{j}(v_{i}+1)(N_{i}-v_{i})(N_{j}-v_{j}+1)\right]
^{1/2}/[N_{i}N_{j}]^{1/2}\nonumber \\
&&\left\langle N_{i},v_{i}-1;N_{j},v_{j}+1\left\vert M_{ij}\right\vert
N_{i},v_{i};N_{j},v_{j}\right\rangle\nonumber\\
&=&-\left[ v_{i}(v_{j}+1)(N_{j}-v_{j})(N_{i}-v_{i}+1)\right]
^{1/2}/[N_{i}N_{j}]^{1/2}. \nonumber
\label{eqn12}
\end{eqnarray}
The Majorana operators have also diagonal matrix elements given by
\begin{eqnarray}
&&\left\langle N_{i},v_{i};N_{j},v_{j}\left\vert M_{ij}\right\vert
N_{i},v_{i};N_{j},v_{j}\right\rangle\\ 
&=&\left(
v_{i}N_{j}+v_{j}N_{i}-2v_{i}v_{j}\right) /\left( N_{i}N_{j}\right) ^{1/2}.
\nonumber
\label{eqn13}
\end{eqnarray}
If the oscillators are equivalent, $N_{i}=N$, their contribution, $\left(
v_{i}+v_{j}-\frac{2v_{i}v_{j}}{N}\right) $ can be reabsorbed into Eq.~(\ref{eqn11})
so it will not be considered further.

The operator
\begin{equation}
\hat M=\sum_{i\ne j}^nM_{ij}
\label{Majorana}
\end{equation}
is Hermitian and its matrix elements are real and symmetric. It is called Majorana (or exchange) operator because it was introduced by Majorana in 1993 in the context of nuclear physics~\cite{Majorana1933}. Within the algebraic theory of molecules~\cite{iac-lev} it was introduced in 1991~\cite{iac-oss2}. Its mathematical definition is that of the invariant operator of $u_1(2)\oplus u_2(2)\oplus\cdots\oplus u_n(2)$. In recent years, operators similar to $\hat M$ have been used in the context of condensed matter physics. 

The elements $M_{ij}$ of the Majorana operator annihilate one quantum of vibration at site $i$ and create one at site $j$, or vice versa. In the harmonic limit, $N_{i}\rightarrow \infty $, the matrix elements of $M_{ij}$ are the familiar matrix elements $\left[ v_{j}(v_{i}+1)\right] ^{1/2}$ of the operator $b_{i}^{\dag }b_{j}$, where $b_{i}^{\dag }$ and $b_{i}$ are boson creation and annihilation operators of a phonon at site $i$, respectively, multiplied by a minus sign. Because of the sum in Eq.~(\ref{Majorana}), $\hat M$ can be written as $\sum_{j>i}^n\left(b_i^\dagger b_j+b_j^\dagger b_i\right)$, which shows explicitly the Hermiticity of $\hat M$. The operator $\hat M$ splits the degeneracy of the states in Fig.~\ref{fig2}, determines the distribution of eigenvalues and thus the level densities for the fundamental vibration $v=\sum_{i}v_{i}=1$ and the overtones $v=2,3,...$, and generates the normal modes of vibrations.

The total Hamiltonian for 1d vibrations of a system of $n$ atoms at
locations $\vec{r}_{i},i=1,...,n$, can be written as%
\begin{equation}
H=E_{0}+\sum_{i=1}^{n}A_{i}C_{i}+\sum_{i\neq j}^{n}A_{ij}^{\prime
}C_{ij}^{\prime }+\sum_{i\neq j}^{n}\lambda _{ij}M_{ij},
\label{eqn14}
\end{equation}%
and for $n$ identical atoms as
\begin{equation}
H=E_{0}+A\sum_{i=1}^{n}C_{i}+A^{\prime }\sum_{i\neq j}^{n}C_{ij}^{\prime
}+\sum_{i\neq j}^{n}\lambda _{ij}M_{ij}.
\label{eqn15}
\end{equation}%
This Hamiltonian applies to any system. In sections~\ref{Chain}, \ref{Square} and~\ref{Hexagon}, we will
consider three cases: (1) linear lattice (1D), (2) square lattice (2D) and
(3) honeycomb lattice (2D).

\section{Linear lattice\label{Chain}}

\begin{figure}[h!]
  {\includegraphics[width=0.6\linewidth]{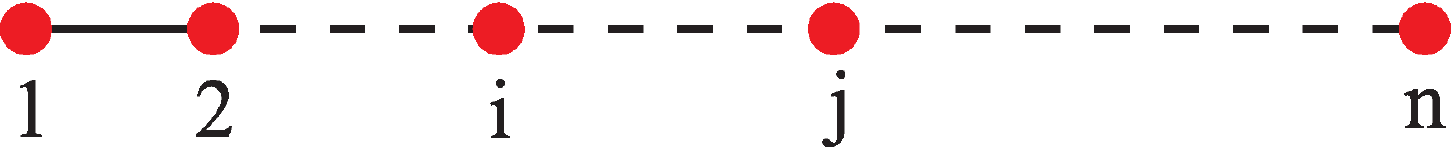}}
  \caption{The linear lattice.}
  \label{fig3}
\end{figure}

The algebraic theory of linear lattices, Fig.~\ref{fig3}, was investigated in \cite{iac-truini}.
The most general algebraic Hamiltonian for this problem can be written as 
\begin{eqnarray}
H &=&A\sum_{j}C_{j}+\lambda ^{(I)}\sum_{j}M_{j,j+1}^{(I)}+\lambda
^{(II)}\sum_{j}M_{j,j+2}^{(II)}\nonumber \\
&+&\lambda ^{(III)}\sum_{j}M_{j,j+3}^{(III)}+...
\label{eqn16}
\end{eqnarray}%
with $j=1,2,...,n$, \ and where $\lambda ^{(I)}$ is the strength of the
nearest-neighbor interaction, $\lambda ^{(II)}$ of the next-nearest
neighbor, $\lambda ^{(III)}$ of the next-to-next one, etc.

The Hamiltonian Eq.~(\ref{eqn16}) is of the general form 
\begin{equation}
H=A\sum_{j}C_{j}+\sum_{j\neq j^{\prime }}\lambda _{jj^{\prime
}}M_{jj^{\prime }}.
\label{eqn17}
\end{equation}%
Denoting by $\left\vert v_{j}\right\rangle $ the number of quanta at each
site $j$, the basis for the diagonalization of Eq.~(\ref{eqn17}) is:
\begin{equation}
\begin{array}{lll}
v=0 &{\rm vacuum} &\left\vert 0\right\rangle =\left\vert 0,0,...,0\right\rangle \\ 
v=1 &n\, {\rm basis\, states} &\left\vert 1_{j}\right\rangle =\left\vert 0,0,...,1_{j},...,0\right\rangle \\ 
v=2 &n\, {\rm basis\, states} 
&\left\vert 2_{j}\right\rangle =\left\vert 0,0,...,2_{j},...,0\right\rangle\\
v=2 &\frac{n(n-1)}{2}\, {\rm basis\, states}
&\left\vert 1_{j},1_{j^{\prime }}\right\rangle =\left\vert
0,0,...,1_{j},...,1_{j^{\prime }},...,0\right\rangle\\
... & ... & ...\\
\end{array}
\label{eqn18}
\end{equation}
The states with $v=1$ constitute the fundamental vibration, the states with $%
v=2$ the overtone and combination modes of the linear lattice. The matrix
elements of the operators $C_{j}$ are 
\begin{eqnarray}
&&\left\langle v_{j}\left\vert C_{j}\right\vert v_{j}\right\rangle \nonumber \\ 
&=&-4\left(v_{j}-\frac{v_{j}^{2}}{N}\right)\nonumber \\
&\overset{N\rightarrow \infty}{\longrightarrow}&-4v_{j}\nonumber \\
&&\left\langle v_{j}+1,v_{j^{\prime }}-1\left\vert M_{jj^{\prime }}\right\vert
v_{j},v_{j^{\prime }}\right\rangle\nonumber \\ 
&=&-\left[ v_{j^{\prime }}(v_{j}+1)(1-%
\frac{v_{j}}{N})(1-\frac{v_{j^{\prime }}-1}{N})\right] ^{1/2}\nonumber \\
&\overset{N\rightarrow \infty}{\longrightarrow}&-\left[ v_{j^{\prime }}(v_{j}+1)\right]^{1/2} .
\label{eqn19}
\end{eqnarray}%
In the harmonic limit, $N\rightarrow \infty $, the algebraic Hamiltonian
Eq.~(\ref{eqn17}) reduces to the boson Hubbard Hamiltonian%
\begin{equation}
H=-4A\sum_{j}b_{j}^{\dag }b_{j}-\sum_{j\neq j^{\prime }}\lambda _{jj^{\prime
}}b_{j}^{\dag }b_{j^{\prime }}
\label{eqn20}
\end{equation}%
where $b_{j}^{\dag }$ creates a quantum at site $j$. For identical
oscillators and nearest-neighbor interaction, it can be written as%
\begin{equation}
H=-4A\sum_{i}\hat{n}_{i}-\lambda \sum_{i}\left( b_{i}^{\dag
}b_{i+1}+b_{i}b_{i+1}^{\dag }\right) ,
\label{eqn21}
\end{equation}%
with, in the notation of Ref.~\cite{kuhner}, $\mu =4A$ and $t=\lambda $.

Since the operator $\sum_{j}C_{j}$ is diagonal in the basis Eq.~(\ref{eqn18}) and the
operator $\sum_{j\neq j^{\prime }}M_{jj^{\prime }}$ conserves the total
number of quanta $v=\sum_{j}v_{j}$, the diagonalization of $H$ splits into
blocks. $H$ can then be diagonalized in the subspaces with $v=1$, $v=2$,
..., of dimension $n,\frac{n(n+1)}{2},...$.

It is of interest to note that the algebraic method can also describe more
complex situations such as those in which the algebraic Hamiltonian is%
\begin{equation}
H=A\sum_{i}C_{i}+B\sum_{i}C_{i}^{2}+B^{\prime }\sum_{i\neq
j}C_{i}C_{j}+\lambda \sum_{i\neq j}M_{ij},
\label{eqn22}
\end{equation}%
which is equivalent to the boson Hubbard Hamiltonian with in-site
interactions%
\begin{eqnarray}
H&=&-\mu \sum_{i}\hat{n}_{i}+U\sum_{i}\hat{n}_{i}(\hat{n}_{i}-1)+V\sum_{i}\hat{%
n}_{i}\hat{n}_{i+1}\nonumber\\
&-&t\sum_{i}\left( b_{i}^{\dag }b_{i+1}+b_{i}b_{i+1}^{\dag}\right) .
\label{eqn23}
\end{eqnarray}%
The diagonalization of this Hamiltonian can be done in the same way as
before since the addidional terms do not modify its block-diagonal form.

\subsection{Solutions\label{ChainSolutions}}

\subsubsection{The fundamental vibration v=1\label{Chainv1}}

\begin{figure}[h!]
{\includegraphics[width=0.8\linewidth]{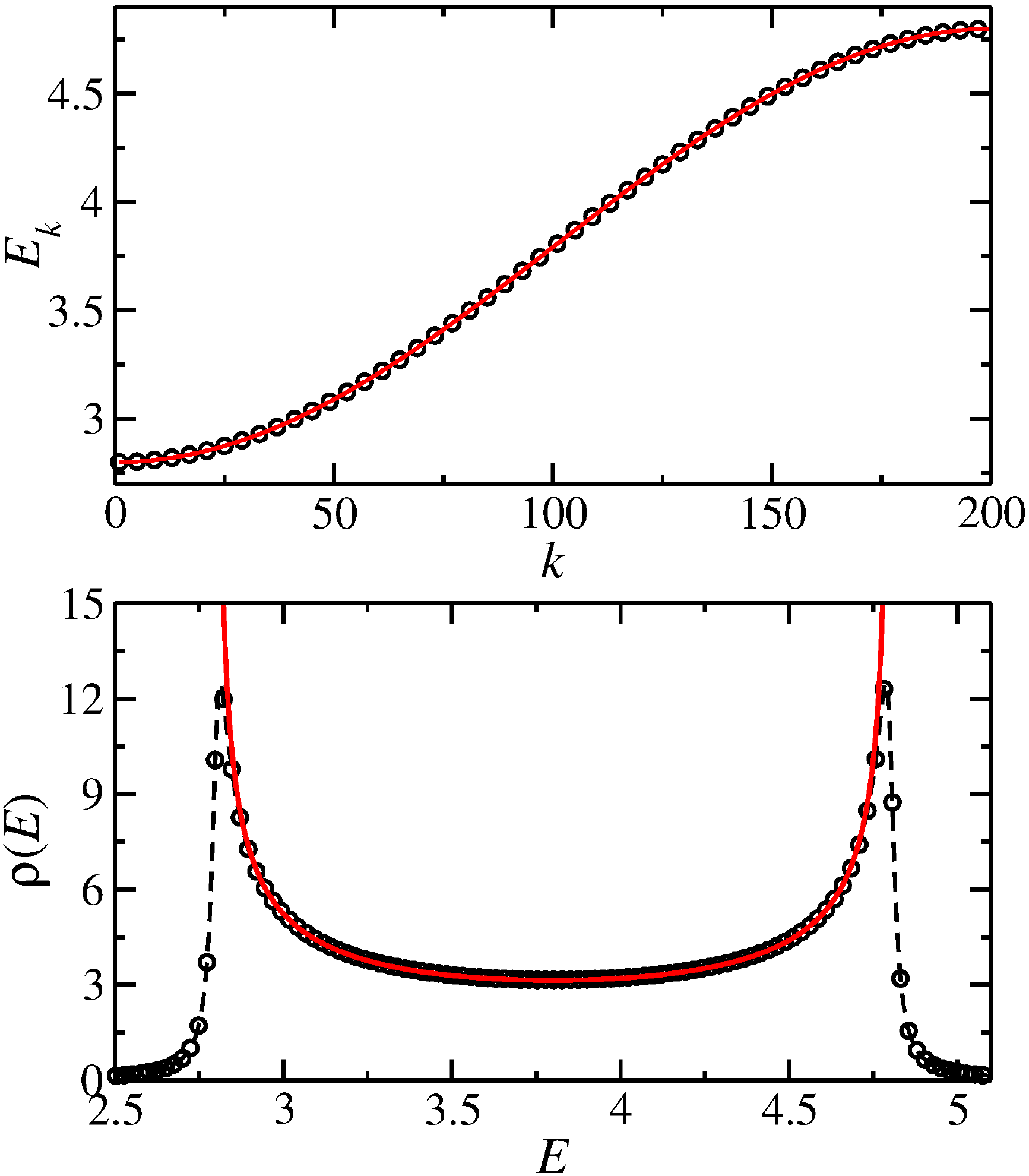}}
\caption{EDR and DOS of the $v=1$ vibration of a linear chain with $v=1,\, \lambda =0.5,\, A=-1,\, n=200,\, N=20000$. Full red lines: Upper panel: Eq.~(\ref{eqn26}), lower panel: Eq.~(\ref{eqn28}).}
\label{fig4}
\end{figure}

The EDR and DOS of the fundamental vibration $v=1$ of a linear lattice can
be obtained analytically by purely algebraic methods by making use of
certain properties of matrices. These properties were first noticed by
Nierenberg \cite{nierenberg} for lines and further exploited and generalized
for lines and rings in Refs.~\cite{iac-truini,iac-del}. The basic property
is that the Majorana matrix on a line $M_{L}$ with matrix elements%
\begin{equation}
\left\langle M_{L}\right\rangle _{t,t\pm s}=c_{s}
\label{eqn24}
\end{equation}%
on the diagonal "$\pm s$ off the main diagonal" has eigenvalues given by 
\begin{equation}
m=c_{s}2\cos \frac{sk\pi }{(n+1)},\, \, \, \, k=1,2,...,n.
\label{eqn25}
\end{equation}%
For nearest-neighbor interactions $s=1$, the solution can be written as%
\begin{equation}
E_{k}=\alpha -2\beta \cos \theta _{k}, \, \, \, \, \theta _{k}=\frac{k\pi}{(n+1)},\, \, \, \, k=1,...,n.
\label{eqn26}
\end{equation}%
This solution applies to both harmonic and anharmonic vibrations. The
coefficients $\alpha $ and $\beta $ are given by%
\begin{equation}
\alpha =-4A\left( 1-\frac{1}{N}\right),\, \, \, \beta =\lambda .
\label{eqn27}
\end{equation}%
If the number of sites is large, one can replace $\frac{k\pi }{(n+1)}$ by a
continuous variable $x$ which runs from $0$ to $\pi $ and $E_{k}$ by a
continuous variable $E$ which runs from $\alpha -2\beta $ to $\alpha +2\beta 
$. The DOS of the $v=1$ vibration is then%
\begin{equation}
\rho (E)=\frac{1}{2\beta \sin \arccos (\frac{\alpha -E}{2\beta })}\, .
\label{eqn28}
\end{equation}%
This result has been verified by numerical simulations as shown in Fig.~\ref{fig4}.
The DOS is symmetric around $E=\alpha $, and has no singularities, except at
the edges $E=\alpha \pm 2\beta $.

\subsubsection{The overtone vibration v=2\label{Chainv2}}

\begin{figure}[h!]
{\includegraphics[width=0.8\linewidth]{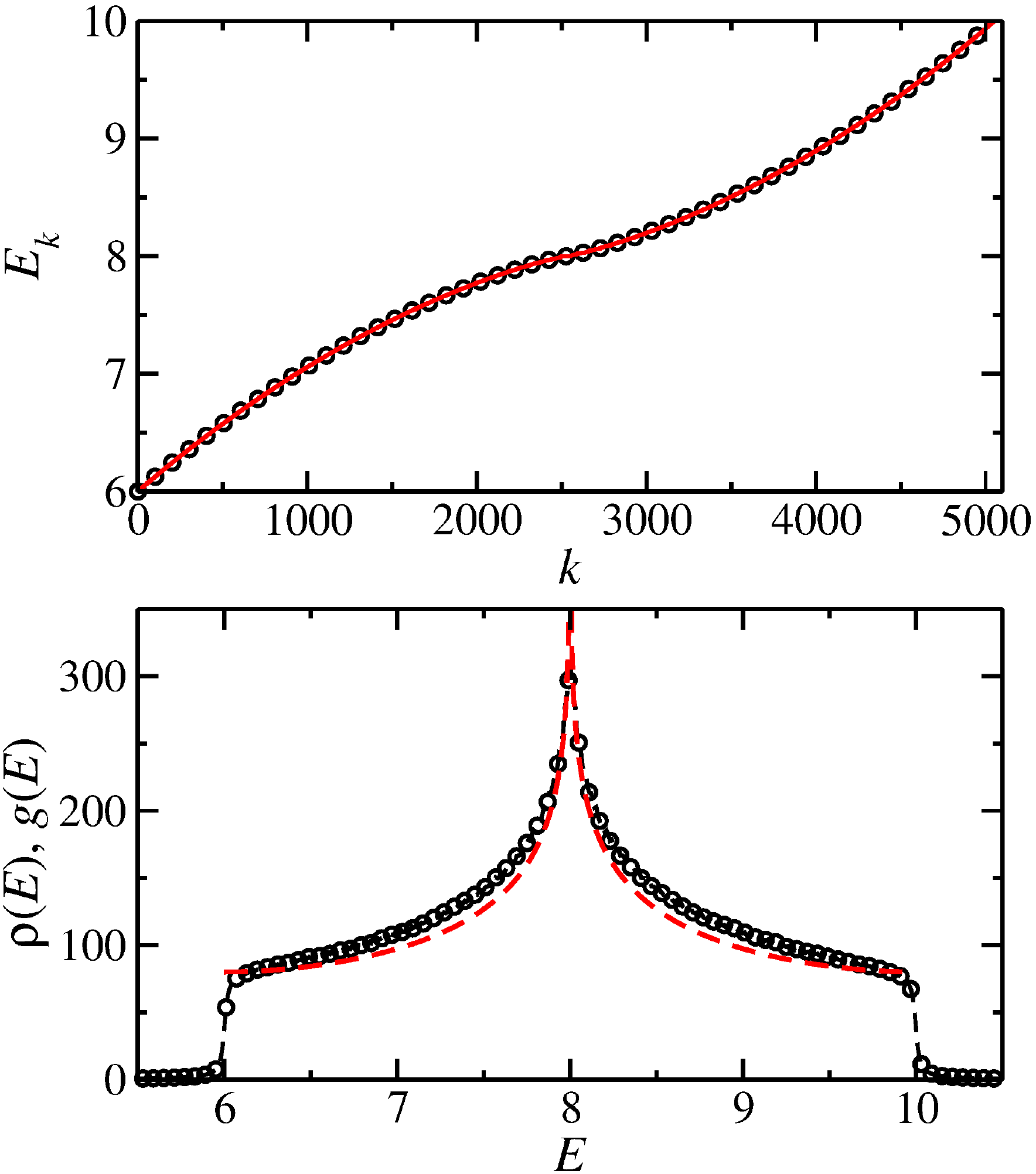}}
\caption{EDR and DOS of the $v=2$ vibration of a linear chain in the harmonic limit with $\lambda =0.5,\, A=-1,\, n=100,\, N=20000$. Full red lines: upper panel: Eq.~(\ref{eqn29}), lower panel: Eq.~(\ref{eqn31}).}
\label{fig5}
\end{figure}

The spectrum and density of states of the overtone vibration can be
calculated analytically only in the harmonic limit. The energies of the $v=2$
vibrations are%
\begin{eqnarray}
E_{k,k^{\prime }}&=&2\alpha -2\beta \left( \cos \theta _{k}+\cos \theta
_{k^{\prime }}\right),\, \, k\leq k^{\prime }, \, \, k,k^{\prime
}=1,...,n;\nonumber\\  
\theta _{k}&=&\frac{k\pi }{(n+1)}.
\label{eqn29}
\end{eqnarray}%
For $n$ large, one can introduce two continuous variables $x=\frac{k\pi }{n+1%
}$ and $y=\frac{k^{\prime }\pi }{n+1}$ and rewrite Eq.~(\ref{eqn29}) as%
\begin{equation}
E(x,y)=2\alpha -2\beta (\cos x+\cos y)
\label{eqn30}
\end{equation}%
with $y\geq x$. The DOS can be calculated from Eq.~(\ref{eqn30}) using
the method introduced by Bowers and Rosenstock \cite{bowers} and discussed
in detail in ~\refsec{Squarev1}, yielding%
\begin{equation}
g(\tilde{E})=\frac{4K(k_{1}^{2})}{\pi ^{2}}, \, \, \, \, k_{1}^{2}=4(1-%
\tilde{E})\tilde{E},
\label{eqn31}
\end{equation}%
where $K$ is a complete elliptic integral of the first kind and $\tilde E=(E-2\alpha +4\beta )/(8\beta )$, $0\leq\tilde E\leq 1$. This DOS has a logarithmic singularity at $\tilde{E}=1/2$. This result is
confirmed by numerical simulations as shown in Fig.~\ref{fig5}
where the energy eigenvalues are plotted as a function of the
single index $k$ which labels the eigenvalues in successive order, instead
of the double indices $k,k^{\prime}$ used in Eq.~(\ref{eqn29}).

\begin{figure}[h!]
  {\includegraphics[width=0.8\linewidth]{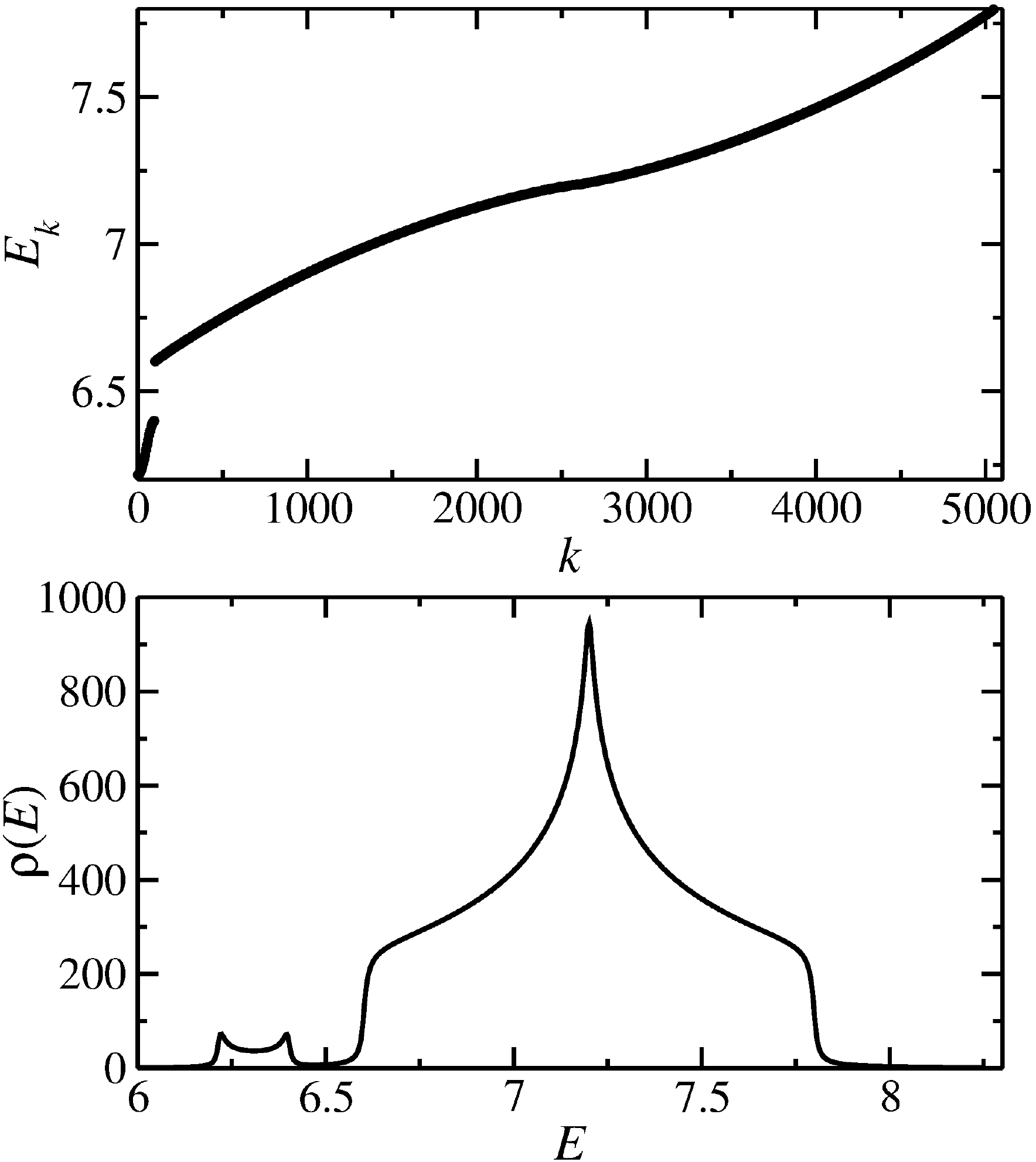}}
  \caption{EDR and DOS of the $v=2$ vibration of a linear anharmonic chain with $\lambda =0.15,\, A=-1,\, n=100,\, N=10$. The EDR exhibits a gap between $E_k=6.4$ and $E_k=6.6$ and, accordingly, the DOS vanishes there.}
  \label{fig6}
\end{figure}

For anharmonic vibrations, analytic solutions are only available for weak $%
\frac{\lambda }{A}\ll 1$ and strong $\frac{\lambda }{A}\gg 1$ couplings \cite%
{iac-truini}. Numerical simulations are shown in Fig.~\ref{fig6}. The EDR exhibits a gap between $E_k=6.4$ and $E_k=6.6$, where the lower sequence contains $n$ eigenvalues and the upper one $n(n-1)/2$, see Eq.~(\ref{eqn18}). The DOS has two edge singularities below the gap and one logarithmic singularity above it.

\section{Square lattice\label{Square}}

For 2D lattices, the DOS depends on the symmetry of the
lattice. We consider first the EDR and DOS of a square lattice.
\begin{figure}[h!]
  {\includegraphics[width=0.27\linewidth]{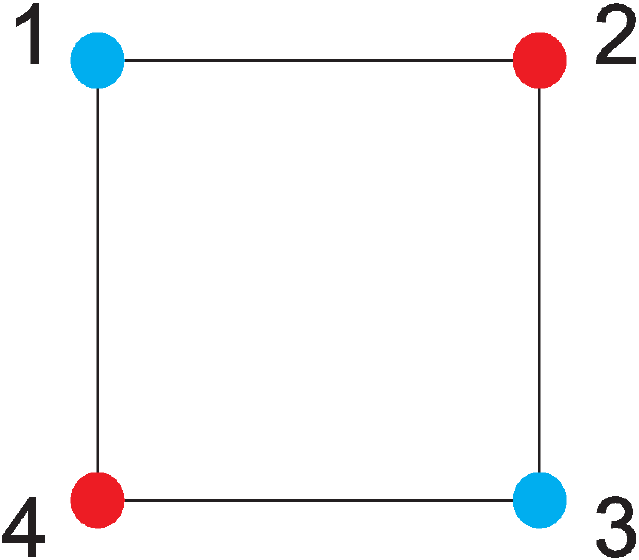}}
  {\includegraphics[width=0.5\linewidth]{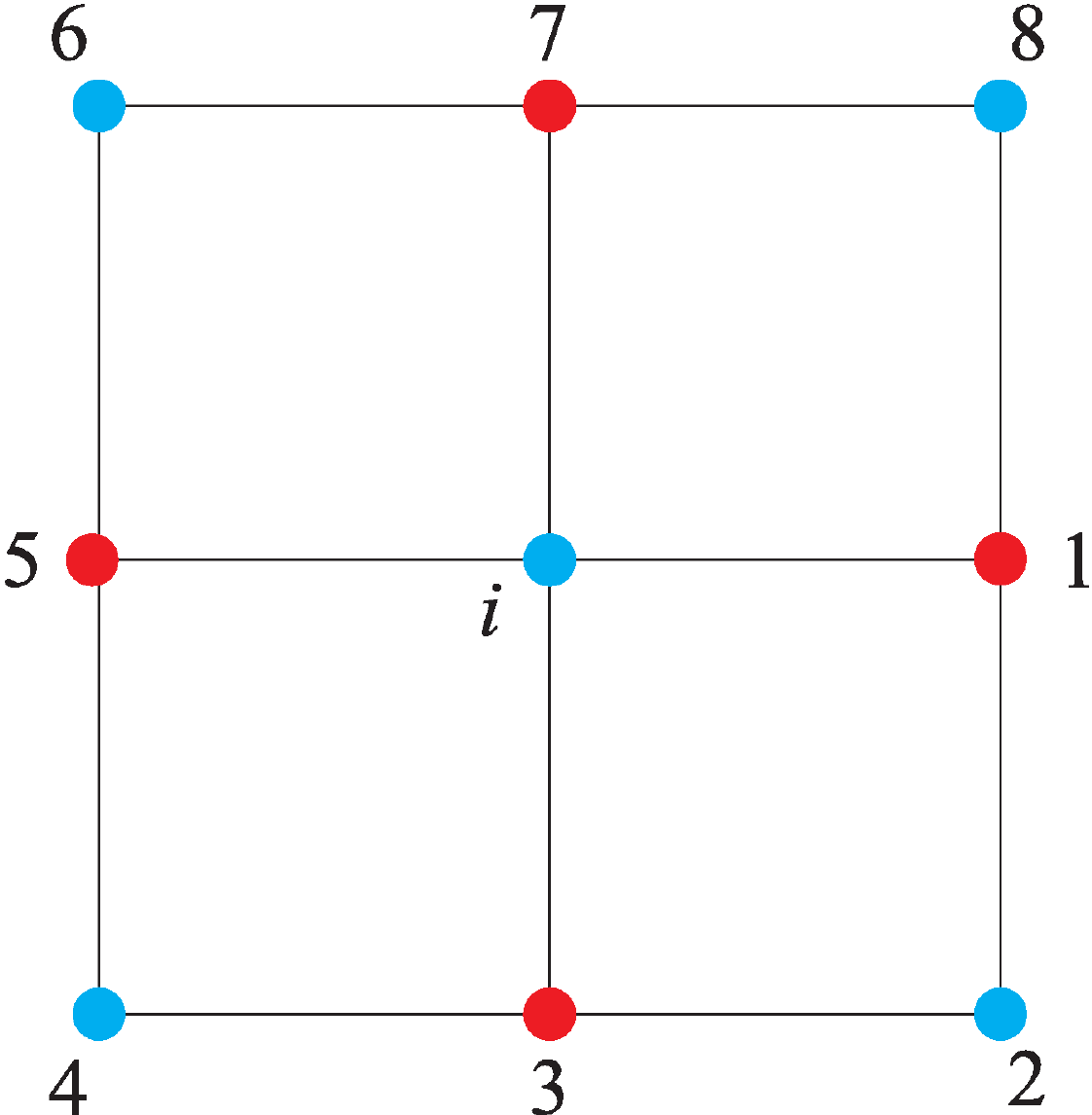}}
  \caption{The unit cell (left) and supercell (right) of a square lattice.}
  \label{fig7}
\end{figure}

A real-space cell of this lattice is shown in Fig.~\ref{fig7} left. In this unit cell
there are two and only two types of interaction, nearest neighbor and
next-to-nearest neighbor. For identical atoms, the symmetry of the unit
cell, $D_{4h}$, imposes the conditions%
\begin{eqnarray}
\lambda ^{(I)} &=&\lambda _{12}=\lambda _{23}=\lambda _{34}=\lambda _{41}, 
\nonumber \\
\lambda ^{(II)} &=&\lambda _{13}=\lambda _{24}.
\label{eqn32}
\end{eqnarray}

A real-space supercell of this lattice is shown in Fig.~\ref{fig7} right. It consists
of four unit cells. The central atom $i$ can interact either with its four
nearest-neighbors 1, 3, 5, 7, or with its four next-to-nearest neighbors
2, 4, 6, 8. The coordinates of the four nearest neighbors are given by%
\begin{equation}
\frac{\boldsymbol{\delta }_{1}}{a}=(1,0),\, \frac{\boldsymbol{\delta }_{3}}{a}=(0,-1),\, \frac{\boldsymbol{\delta }_{5}}{a}=(-1,0),\, \frac{\boldsymbol{\delta }_{7}}{a}=(0,1),
\label{eqn33}
\end{equation}%
while those of the next-to-nearest neighbors are%
\begin{equation}
\frac{\boldsymbol{\delta }_{2}}{a}=(1,1),\, \frac{\boldsymbol{\delta }_{4}}{a}=(1,-1),\, %
\frac{\boldsymbol{\delta }_{6}}{a}=(-1,-1),\, \frac{\boldsymbol{\delta }_{8}}{a}=(-1,1),
\label{eqn34}
\end{equation}%
where $a$ is the lattice constant. The symmetry of the lattice imposes 
the conditions%
\begin{eqnarray}
\lambda ^{(I)} &=&\lambda _{i1}=\lambda _{i3}=\lambda _{i5}=\lambda _{i7},
\nonumber \\
\lambda ^{(II)} &=&\lambda _{i2}=\lambda _{i4}=\lambda _{i6}=\lambda _{i8}.
\label{eqn35}
\end{eqnarray}%
The Majorana interaction can then be written as%
\begin{equation}
M=\lambda ^{(I)}\sum_{\left\langle i,j\right\rangle }M_{ij}^{(I)}+\lambda
^{(II)}\sum_{\left\langle \left\langle i,j\right\rangle \right\rangle
}M_{ij}^{(II)}.
\label{eqn36}
\end{equation}%
In the harmonic limit%
\begin{equation}
M=\lambda ^{(I)}\sum_{\left\langle i,j\right\rangle }\left( b_{i}^{\dagger
}b_{j}+h.c.\right) +\lambda ^{(II)}\sum_{\left\langle \left\langle
i,j\right\rangle \right\rangle }\left( b_{i}^{\dagger }b_{j}+h.c.\right) .
\label{eqn37}
\end{equation}%
The full Hamiltonian for $n$ atoms in a square lattice is 
\begin{eqnarray}
H&=&E_{0}+A\sum_{i}C_{i}+A^{\prime }\sum_{i<j}C_{ij}^{\prime }+\lambda
^{(I)}\sum_{\left\langle i,j\right\rangle }M_{i,j}^{(I)}\nonumber\\
&+&\lambda^{(II)}\sum_{\left\langle \left\langle i,j\right\rangle \right\rangle
}M_{i,j}^{(II)}.
\label{eqn38}
\end{eqnarray}%
The operators%
\begin{equation}
\sum_{\left\langle i,j\right\rangle }M_{ij}^{(I)}\equiv S^{(I)},\, \, 
\sum_{\left\langle \left\langle i,j\right\rangle \right\rangle
}M_{ij}^{(II)}\equiv S^{(II)},
\label{eqn39}
\end{equation}%
are also called symmetry adapter operators.

\subsection{Solutions\label{SquareSolutions}}

\subsubsection{The fundamental vibration v=1\label{Squarev1}}

The EDR and the DOS of the fundamental vibration $%
v=1$ of a square lattice with nearest-neighbor and next-to-nearest neighbor
interactions and Hamiltonian (38) can be obtained analytically by purely
algebraic methods by making use of certain properties of matrices as
discussed in~\refsec{Chainv1}, Eqs.~(\ref{eqn24}) and~(\ref{eqn25}).

The Hamiltonian matrix is obtained by taking matrix elements of $H$,
Eq.~(\ref{eqn38}), in the basis $\left\vert 1,0,0,...\right\rangle ,\left\vert
0,1,0,...\right\rangle ,...,\left\vert 0,0,...,0,1\right\rangle \equiv
\left\vert 1_{j}\right\rangle $ of Eq.~(\ref{eqn18}). The operators $C_{i}$, $%
C_{ij}^{\prime }$ are diagonal in this basis with values given by $-4(1-%
\frac{1}{N})$ and $-4(\frac{1}{2N})$, Eq.~(\ref{eqn11}). The interesting operators
are $S^{(I)}$ and $S^{(II)}$. The matrices representative of these operators
are of the type discussed previously and thus can be diagonalized
analytically. The solution for the fundamental vibration of a square lattice
was first given by Bowers and Rosenstock \cite{bowers} in terms of functions
over an area and subsequently by Nierenberg \cite{nierenberg} in terms of
functions over a line. In Ref.~\cite{bowers} the eigenvalues of $%
S^{(I)}$ and $S^{(II)}$ are labelled by two integers $p,q=1,2,...,n$ and
given by 
\begin{eqnarray}
m^{(I)}(p,q) &=&2\left( \cos \frac{p\pi }{n+1}+\cos \frac{q\pi }{n+1}\right)
\nonumber \\
m^{(II)}(p,q) &=&4\left( \cos \frac{p\pi }{n+1}\cos \frac{q\pi }{n+1}\right)
.
\label{eqn40}
\end{eqnarray}%
From these, one can construct the eigenvalues of the Majorana operator $%
M=\lambda ^{(I)}S^{(I)}+\lambda ^{(II)}S^{(II)}$ as%
\begin{eqnarray}
m(p,q)&=&\lambda ^{(I)}2\left( -\cos \frac{p\pi }{n+1}-\cos \frac{q\pi }{n+1}%
\right)\nonumber\\
&+&\lambda ^{(II)}4\left( -\cos \frac{p\pi }{n+1}\cos \frac{q\pi }{n+1}%
\right) ,
\label{eqn41}
\end{eqnarray}%
where the minus sign has been introduced to conform with the definition in
Eq.~(\ref{eqn12}). These solutions can be easily verified for the unit cell, $n=4$,
and for the supercell, $n=9$. For the unit cell, $n=4$, the eigenvalues are
given in the terms of $\cos (\frac{p\pi }{3})$ and are%
\begin{eqnarray}
(I) &:&\, \, \pm 2,0,0  \nonumber \\
(II)&:&\, \, \pm 1,\pm 1
\label{eqn42}
\end{eqnarray}%
For the supercell, $n=9$, they are 
\begin{eqnarray}
(I) &:&\, \, \pm 2\sqrt{2},\pm \sqrt{2},\pm \sqrt{2},0,0,0  \nonumber
\\
(II) &:&\, \, \pm 2,\pm 2,0,0,0,0,0
\label{eqn43}
\end{eqnarray}%
For the unit cell the eigenstates are representations of $D_{4h}$. The
eigenstates of (I) are the singly degenerate representations $A$, $B$ and
the doubly degenerate representation $E$~\cite{Wilson1955}.

When $n$ is very large, one may replace $\frac{p\pi }{n+1}$ and $\frac{q\pi 
}{n+1}$ by continuous variables $x$ and $y$, respectively, which run from $0$
to $\pi $,%
\begin{eqnarray}
m(x,y)&=&\lambda ^{(I)}2\left( -\cos x-\cos y\right)\nonumber\\ 
&+&\lambda ^{(II)}4\left(-\cos x\cos y\right),\, \, \, 0\leq x,y\leq \pi .
\label{eqn44}
\end{eqnarray}%
The DOS can be obtained from Eq.~(\ref{eqn44}). For purposes of display, it
is convenient to consider the shifted function \cite{bowers}%
\begin{eqnarray}
f(x,y)&=&\lambda ^{(I)}2\left( 2-\cos x-\cos y\right)\nonumber\\ 
&+&\lambda ^{(II)}4\left( 1-\cos x\cos y\right) ,\, \, 0\leq x,y\leq \pi .
\label{eqn45}
\end{eqnarray}%
The DOS $g(E)$ is obtained from Eq.~(\ref{eqn44}) by 
\begin{equation}
g(E)=\frac{1}{\pi^{2}}\int_{0}^{\pi}\int_{0}^{\pi}\delta \left( E-f(x,y)\right) dxdy.
\label{eqn46}
\end{equation}
\begin{figure}[h!]
{\includegraphics[width=\linewidth]{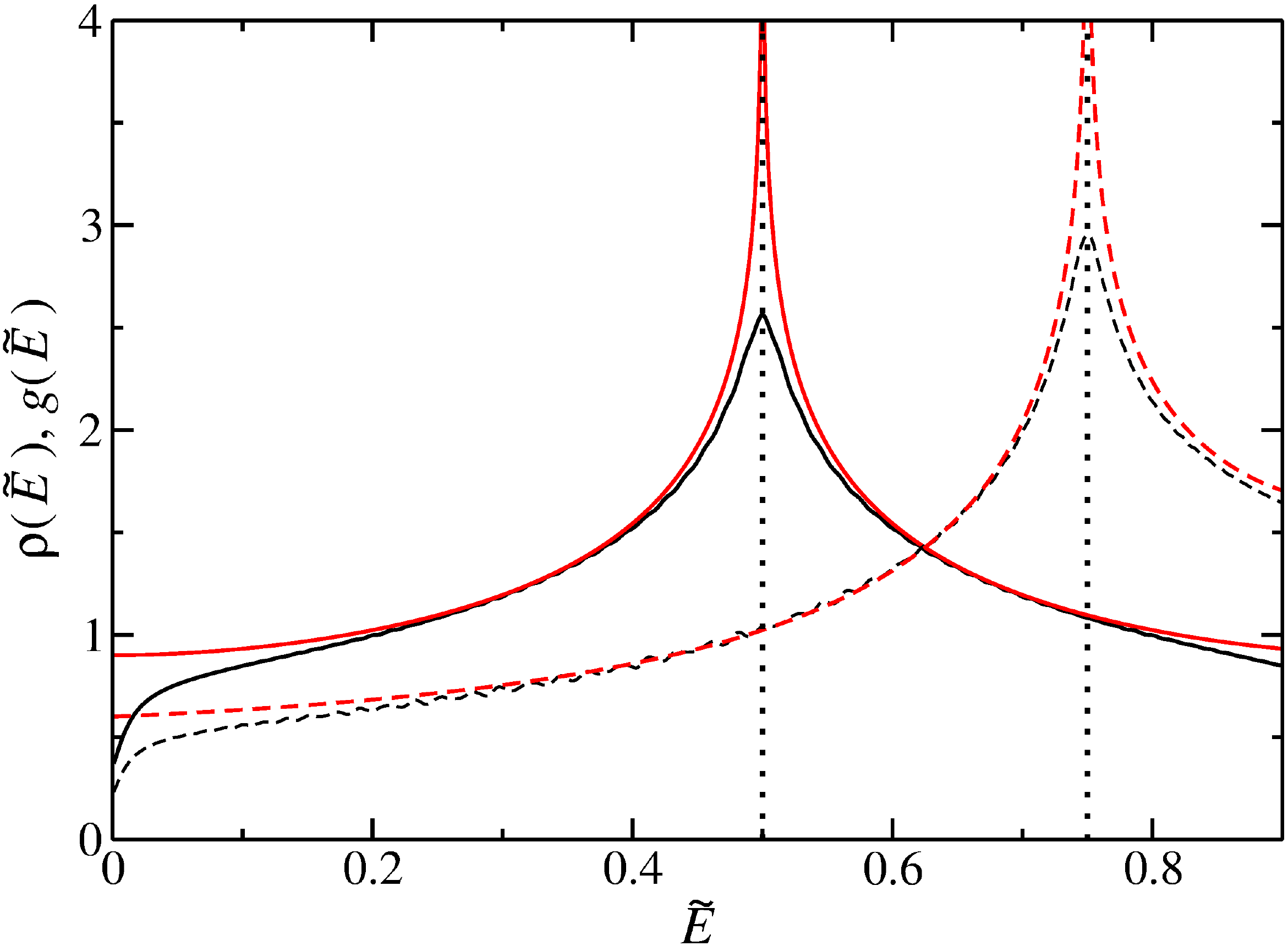}}
\caption{Black: DOS of the fundamental vibration $v=1$ of a square lattice with $n=2500,\, A=-1,\, N=20000$. Full lines: $\lambda^{(II)}/\lambda^{(I)}=0$. Dashed lines: $\lambda^{(II)}/\lambda^{(I)}=0.25$. Red: Eq.~(\ref{eqn47}).}
\label{fig8}
\end{figure}
The DOS is particularly simple when $\lambda ^{(II)}/\lambda ^{(I)}\leq 
\frac{1}{2}$. In this case, introducing \ $E_{\max }=8\lambda ^{(I)}$ and $%
\tilde{E}=E/E_{\max }$, it can be written in terms of an elliptic integral,%
\begin{eqnarray}
g(\tilde{E}) &=&\frac{4K(k_{1}^{2})}{\pi ^{2}\sqrt{\left( 1+2\lambda
^{(II)}/\lambda ^{(I)}\right) ^{2}-8\left( \lambda ^{(II)}/\lambda
^{(I)}\right)\tilde{E}}},\\
k_{1}^{2} &=&\frac{4\left( 1-\tilde{E}\right) \tilde{E}}{\left( 1+2\lambda
^{(II)}/\lambda ^{(I)}\right) ^{2}-8\left( \lambda ^{(II)}/\lambda
^{(I)}\right) ^{2}\tilde{E}},\, \, \, \, \, 0<\tilde{E}<1.\nonumber
\label{eqn47}
\end{eqnarray}%
For nearest-neighbor interactions, $\lambda ^{(II)}=0$, the DOS takes the
form%
\begin{equation}
g(\tilde{E})=\frac{4K(k_{1}^{2})}{\pi ^{2}},\, \, \, k_{1}^{2}=4(1-%
\tilde{E})\tilde{E},
\label{eqn48}
\end{equation}%
which is symmetric around $\tilde{E}=1/2$. The DOS $g(\tilde{E})$ is
shown in Fig.~\ref{fig8}, for purely nearest-neighbor interactions, $\lambda ^{(II)}=0$%
, and next-to-nearest neighbor interactions, with $\lambda ^{(II)}=\frac{1}{4%
}\lambda ^{(I)}$.

\begin{figure}[h!]
{\includegraphics[width=0.8\linewidth]{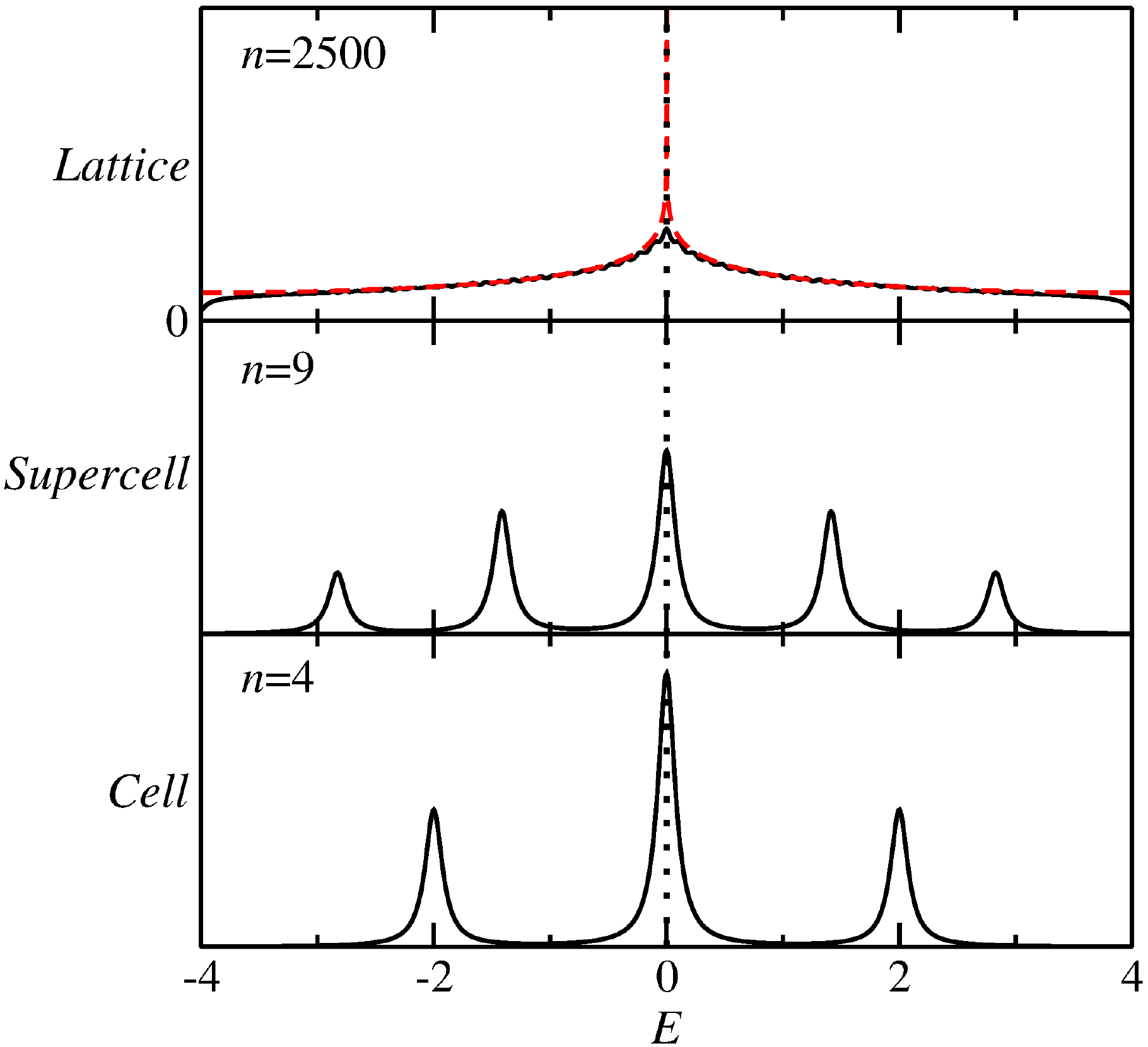}}
\caption{DOS of the fundamental vibration $v=1$ of a square lattice including the unit cell and the supercell with the number of sites $n$ as indicated in the insets. Red dashed line: Eq.~(\ref{eqn47}).}
\label{fig9}
\end{figure}

From this "reduced" DOS, one can reconstruct the "true" DOS. The energy of
the $v=1$ states including the diagonal terms is given by%
\begin{equation}
E(p,q)=E_{0}-4A\left( 1-\frac{1}{N}\right) -4A^{\prime }(\frac{1}{2N}%
)+m(p,q).
\label{eqn49}
\end{equation}%
The "true"\ DOS is obtained from Eq.~(\ref{eqn45}) by changing $\tilde{E}$ to $E=8\lambda^{(I)}\tilde{E}$ and shifting it by the amount $E_{0}-4A\left( 1-\frac{1}{N}%
\right) -4A^{\prime }\left( \frac{1}{2N}\right) -4\lambda ^{(I)}-4\lambda
^{(II)}$. It is interesting to note that the structure of the DOS is a
consequence of the symmetry $D_{4h}$ of the unit cell, and is already
encoded into the DOS of the unit cell and the supercell, as shown in Fig.~\ref{fig9}.

\begin{figure}[h!]
{\includegraphics[width=0.8\linewidth]{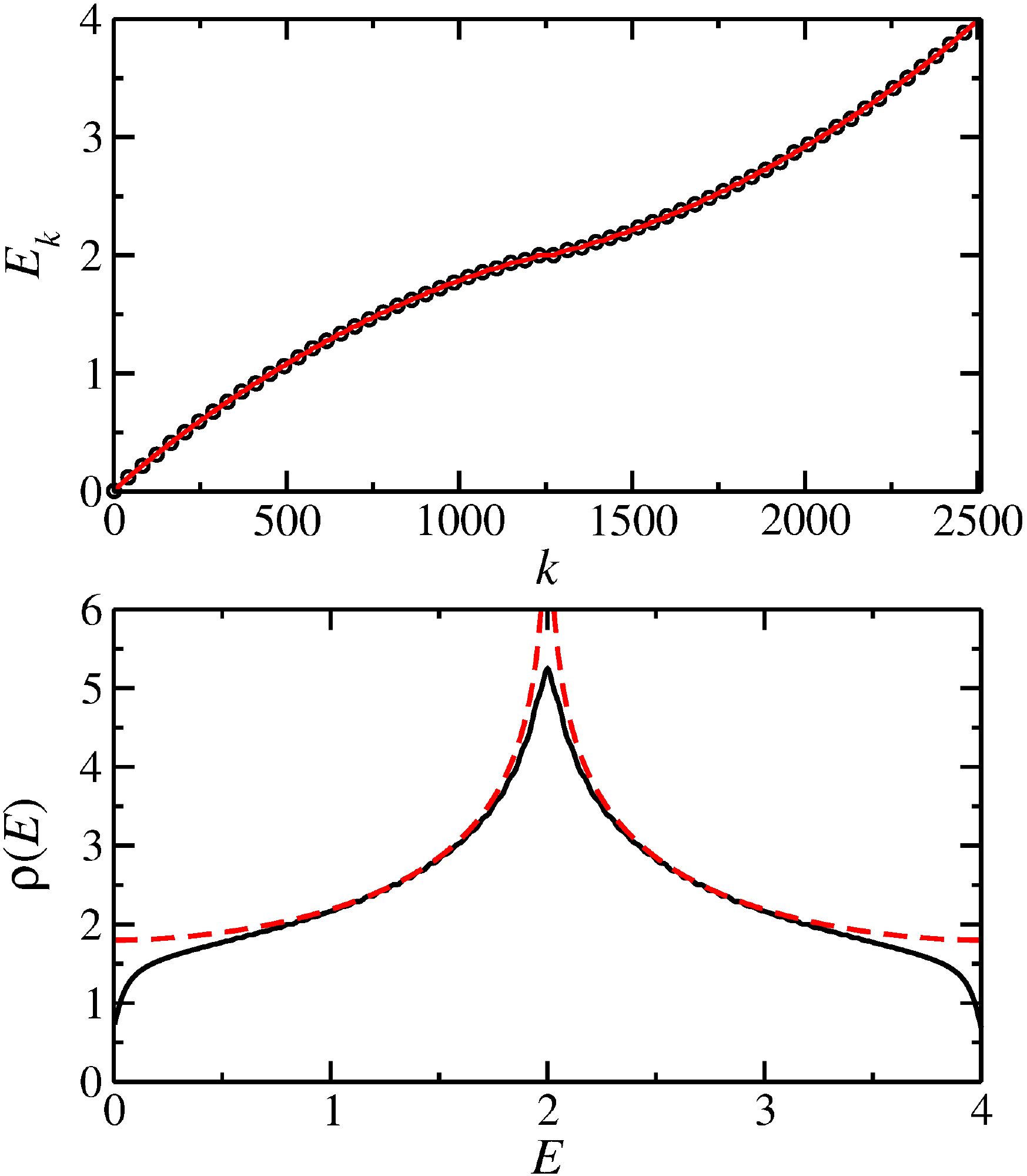}}
\caption{EDR and DOS of the fundamental vibration $v=1$ of a harmonic square lattice with $n=2500,\, A=-1,\, N=20000,\, \lambda^{(I)}=0.5,\, \lambda^{(II)}=0$. Red lines:  Eq.~(\ref{eqn47}).}
\label{fig10}
\end{figure}

This result has been verified by numerical simulations as shown in Fig.~\ref{fig10}. The DOS of the $v=1$ vibration of a square lattice has a logarithmic
singularity at $\tilde{E}=1/2$.

It should be noted that the DOS of the fundamental vibration, $v=1$, of the
2D lattice is identical to that of the overtone vibration $v=2$ of the 1D
lattice. This result follows from the mapping of the function $f(x,y)$
defined over an area onto a function defined over a line, as discussed by
Nierenberg \cite{nierenberg}.

\subsubsection{The overtone v=2\label{Squarev2}}

While for the fundamental vibration $v=1$ the EDR and the DOS do not depend on whether or not the crystal is harmonic, for the
overtone $v=2$ they do. The spectrum of states can be calculated by
diagonalizing the Hamiltonian $H$ in the basis $\left\vert
2,0,0,...\right\rangle $, $\left\vert 1,1,0,...\right\rangle $, $\left\vert
1,0,1,...\right\rangle $, $...$ of Eq.~(\ref{eqn18}), with in total two quanta, in terms
of the four parameters, $A$, $A^{\prime }$, $\lambda ^{(I)}$, $\lambda
^{(II)}$ and the anharmonicity parameter $N$. Analytic solutions can be
obtained only in the weak coupling $\left( \lambda /A\ll 1\right) $ and
strong coupling limit $\left( \lambda /A\gg 1\right)$ \cite{iac-del,iac-truini}.
\begin{figure}[h!]
{\includegraphics[width=0.8\linewidth]{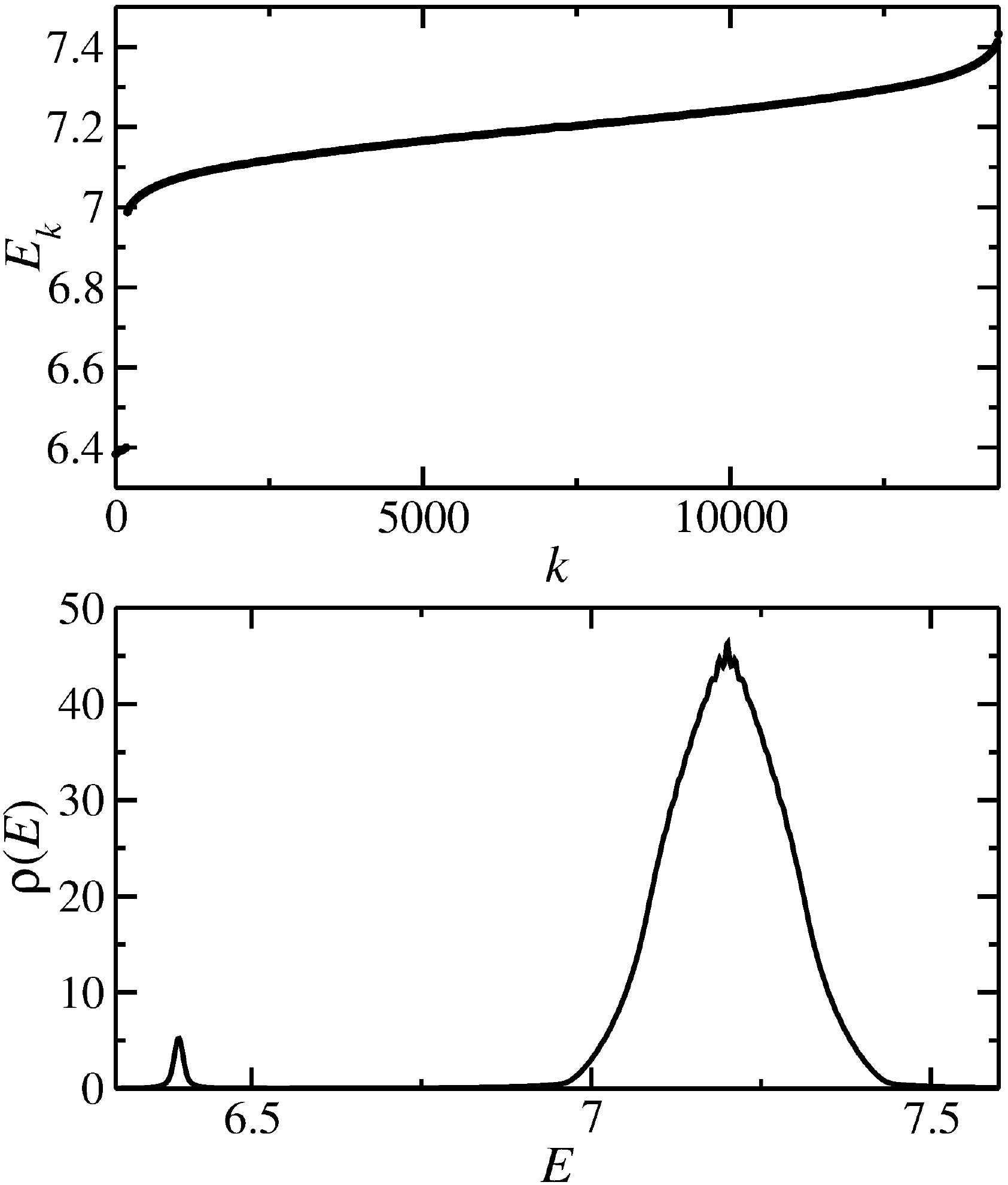}}
\caption{EDR and DOS of the $v=2$ overtone of an anharmonic square lattice with $n=169,\, A=-1,\, N=10,\, \lambda=0.03$.}
\label{fig11}
\end{figure}

\begin{figure}[h!]
{\includegraphics[width=0.8\linewidth]{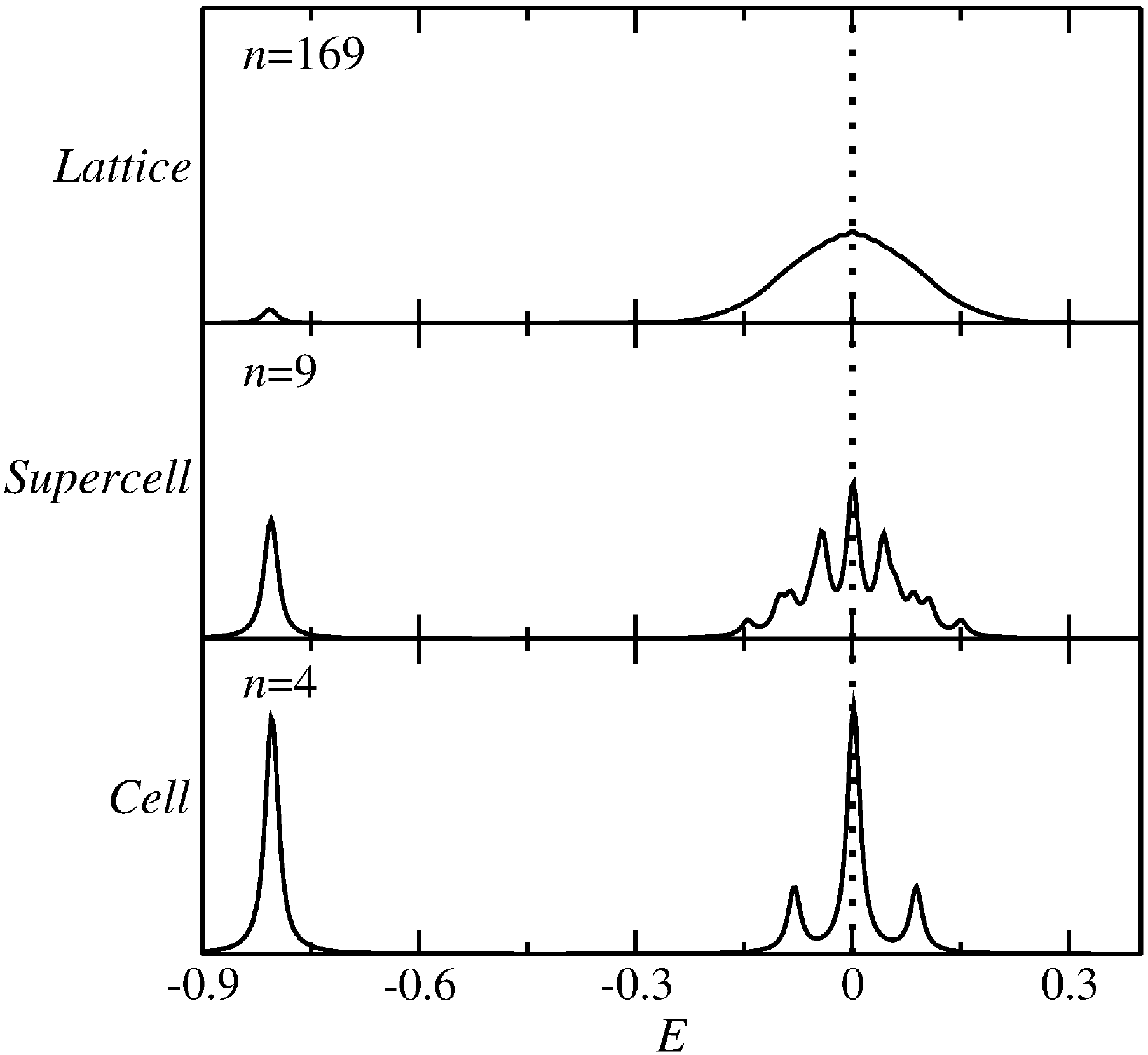}}
\caption{DOS of the $v=2$ overtone of an anharmonic square lattice including the unit cell and the supercell with the number of sites indicated in the insets and $A=-1,\, N=10,\, \lambda=0.03$.}
\label{fig12}
\end{figure}

In order to illustrate the properties of the solutions, we consider the case
of a highly anharmonic vibration with $N=10,$ $-4A=4$, $-4A^{\prime }=0$,
and only nearest-neighbor interactions $\lambda ^{(I)}=0.03$, $\lambda
^{(II)}=0$. The DOS of $v=2$ states has two parts. Both parts diverge as $%
n\rightarrow \infty $, the first part as $n$ and the second as $n^{2}$. This
is seen in the numerical simulation in Fig.~\ref{fig11}. The splitting into two pieces containing $n$ and $n(n-1)/2$ eigenvalues, respectively,
can already be seen in the unit cell, $n=4$, and in the supercell, $n=9$, as
shown in Fig.~\ref{fig12}.

\section{Honeycomb lattice\label{Hexagon}}

\begin{figure}[h!]
{\includegraphics[width=0.3\linewidth]{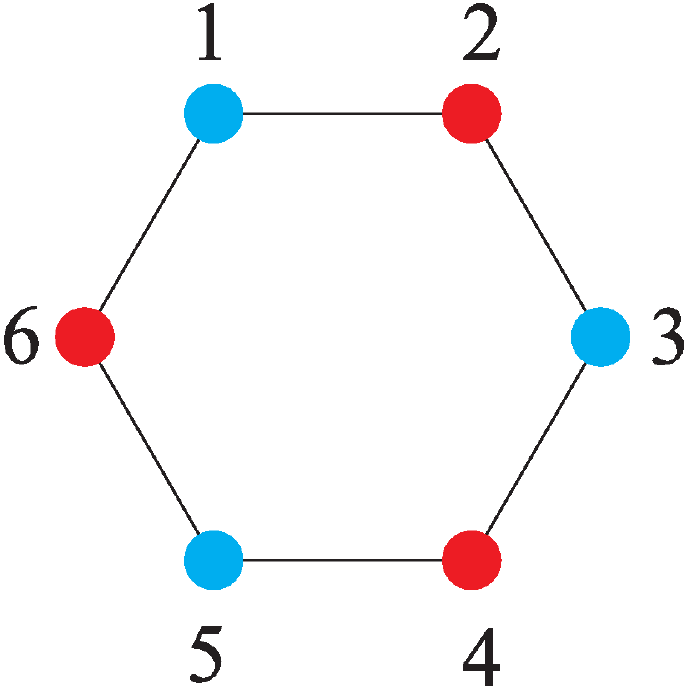}}
%\psfrag{d1}[cc][cc]{$\boldsymbol{\delta}_1$}
%\psfrag{d2}[cc][cc]{$\boldsymbol{\delta}_2$}
%\psfrag{d3}[cc][cc]{$\boldsymbol{\delta}_3$}
{\includegraphics[width=0.5\linewidth]{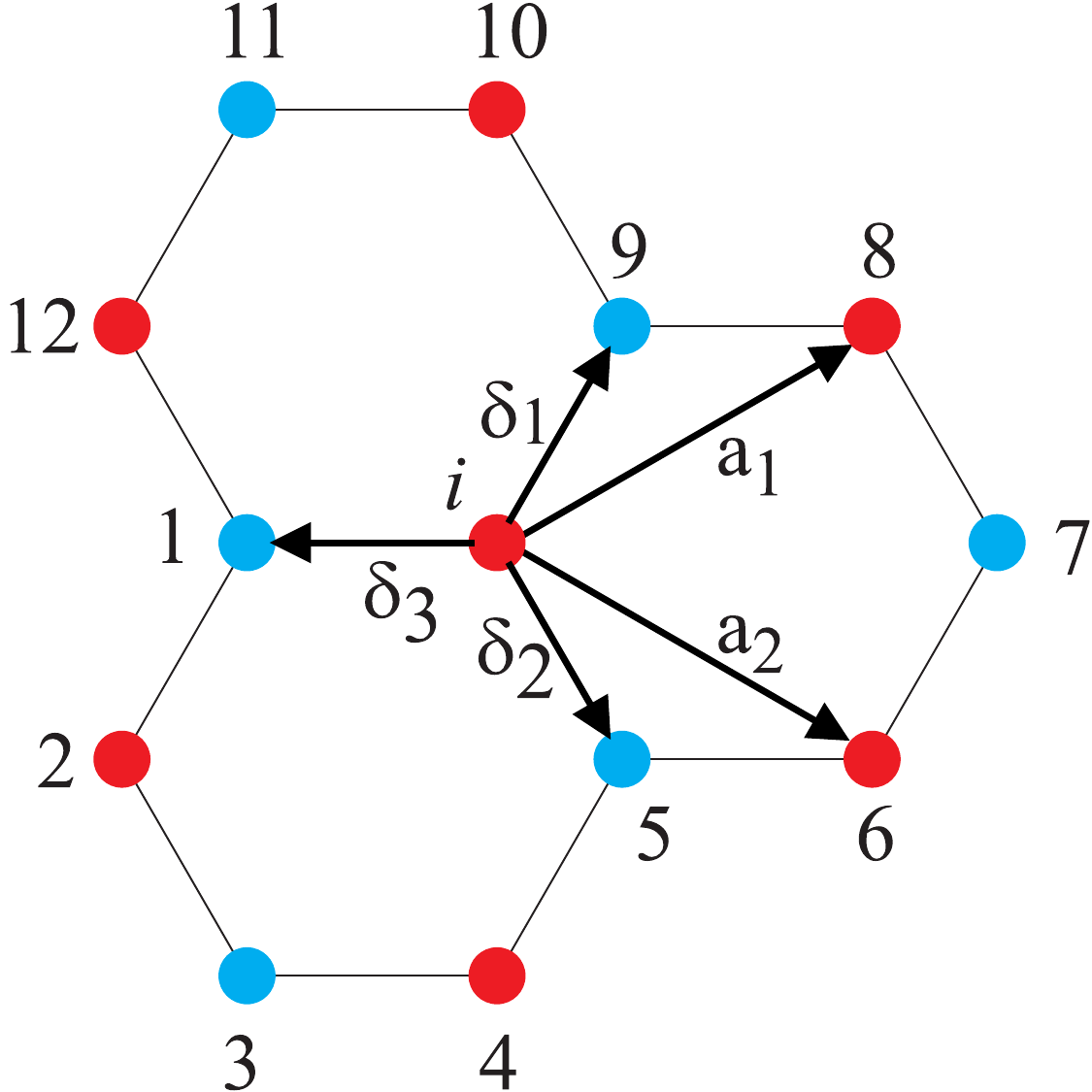}}
  \caption{The unit cell of the honeycomb lattice, (left), and its
supercell, (right).}
  \label{fig13}
\end{figure}

A real-space cell of this lattice is shown in the left part of Fig.~\ref{fig13}.
In this unit cell there are three and only three types of interaction, (I)
nearest neighbor, (II) next-to-nearest neighbor, and (III) third neighbor.
For identical atoms the symmetry of the unit cell, $D_{6h}$, imposes the
condition~(\cite{iac-lev}, page 139) for the interactions,%
\begin{eqnarray}
\lambda ^{(I)} &=&\lambda _{12}=\lambda _{23}=\lambda _{34}=\lambda
_{45}=\lambda _{56}=\lambda _{16},  \nonumber \\
\lambda ^{(II)} &=&\lambda _{13}=\lambda _{24}=\lambda _{35}=\lambda
_{46}=\lambda _{15}=\lambda _{26},  \nonumber \\
\lambda ^{(III)} &=&\lambda _{14}=\lambda _{25}=\lambda _{36}.
\label{eqn50}
\end{eqnarray}

A real-space supercell of this lattice is shown in Fig.~\ref{fig13} right. It consists
of three unit hexagonal cells. The central atom $i$ can interact either with
its three nearest neighbors 1,5,9, or with its six next-to-nearest neighbors
2,4,6,8,10,12, or with its three third neighbors 3,7,10. The coordinates of
the nearest neighbors are given by%
\begin{equation}
\boldsymbol{\delta }_{1}\boldsymbol{=}\frac{a}{2}\left( 1,\sqrt{3}\right) ,\,
\, \boldsymbol{\delta }_{2}=\frac{a}{2}\left( 1,-\sqrt{3}\right) ,\, \, 
\boldsymbol{\delta }_{3}=-a\left( 1,0\right) ,
\label{eqn51}
\end{equation}%
while those of the next-to-nearest neighbor, $\boldsymbol{\delta }^{\prime }$,
and third neighbor, $\boldsymbol{\delta }"$, are given in terms of the lattice
vectors%
\begin{equation}
\boldsymbol{a}_{1}=\frac{a}{2}\left( 3,\sqrt{3}\right) ,\, \, \, \boldsymbol{a}%
_{2}=\frac{a}{2}\left( 3,-\sqrt{3}\right) ,
\label{eqn52}
\end{equation}%
as%
\begin{eqnarray}
\boldsymbol{\delta }_{1}^{\prime } &=&\pm \boldsymbol{a}_{1},\, \, \boldsymbol{%
\delta }_{2}^{\prime }=\pm \boldsymbol{a}_{2},\, \, \, \boldsymbol{\delta }%
_{3}^{\prime }=\pm \left( \boldsymbol{a}_{2}-\boldsymbol{a}_{1}\right)\\
\boldsymbol{\delta }_{1}" &=&\boldsymbol{a}_{1}+\boldsymbol{\delta }_{2},\, \, 
\boldsymbol{\delta }_{2}"=-\boldsymbol{a}_{1}+\boldsymbol{\delta }_{2},\, \, 
\boldsymbol{\delta }_{3}"=-\boldsymbol{a}_{2}+\boldsymbol{\delta }_{1}.\nonumber
\label{eqn53}
\end{eqnarray}%
The symmetry of the unit cell of the honeycomb lattice, $D_{6h}$, imposes
the conditions%
\begin{eqnarray}
\lambda ^{(I)} &=&\lambda _{i1}=\lambda _{i5}=\lambda _{i9},  \nonumber \\
\lambda ^{(II)} &=&\lambda _{i2}=\lambda _{i4}=\lambda _{i6}=\lambda
_{i8}=\lambda _{i10}=\lambda _{i12},  \nonumber \\
\lambda ^{(III)} &=&\lambda _{i3}=\lambda _{i7}=\lambda _{i10}.
\label{eqn54}
\end{eqnarray}%
The Majorana interaction for the honeycomb lattice can be written as 
\begin{eqnarray}
M &=&\lambda ^{(I)}\sum_{\left\langle i,j\right\rangle }M_{ij}^{(I)}+\lambda
^{(II)}\sum_{\left\langle \left\langle i,j\right\rangle \right\rangle
}M_{ij}^{(II)}  \nonumber \\
&+&\lambda ^{(III)}\sum_{\left\langle \left\langle \left\langle
i,j\right\rangle \right\rangle \right\rangle }M_{ij}^{(III)}.
\label{eqn55}
\end{eqnarray}%
In the harmonic limit, $M$ becomes the Hamiltonian of the tight-binding model~\cite{castro}%
\begin{eqnarray}
M &=&\lambda ^{(I)}\sum_{\left\langle i,j\right\rangle }\left( b_{i}^{\dag
}b_{j}+h.c.\right) +\lambda ^{(II)}\sum_{\left\langle \left\langle
i,j\right\rangle \right\rangle }\left( b_{i}^{\dag }b_{j}+h.c.\right)  
\nonumber \\
&+&\lambda ^{(III)}\sum_{\left\langle \left\langle \left\langle
i,j\right\rangle \right\rangle \right\rangle }\left( b_{i}^{\dag
}b_{j}+h.c.\right) .
\label{eqn56}
\end{eqnarray}

The honeycomb lattice can also be viewed as two interpenetrating triangular
lattices A and B \cite{wallace}. In the unit cell shown in the left part of Fig.~\ref{fig13}, the three atoms 1, 5, 9 belong to A and the three atoms 2, 4, 6 belong to B. Neglecting third neighbor
interactions, denoting by $a_{i}^{\dag }$ and $b_{i}^{\dag }$ the boson
creation operators at site $i$ on sublattices A and B, Eq.~(\ref{eqn41}) can be
rewritten as 
\begin{equation}
M=-t\sum_{\left\langle i,j\right\rangle }\left( a_{i}^{\dag
}b_{j}+h.c.\right) -t^{\prime }\sum_{\left\langle \left\langle
i,j\right\rangle \right\rangle }\left( a_{i}^{\dag }a_{j}+b_{i}^{\dag
}b_{j}+h.c.\right) ,
\label{eqn57}
\end{equation}%
where $t=\lambda ^{(I)}$ and $t^{\prime }=\lambda ^{(II)}$.

The Majorana operator in Eq.~(\ref{eqn56}) is written in terms of boson operators. Its
structure, however, depends only on the symmetry of the lattice. The
symmetry adaptation in Eq.~(\ref{eqn56}) can therefore also be used for fermions.
Introducing fermion creation operators $a_{i,\sigma }^{\dag },b_{i,\sigma
}^{\dag }$, the Hamiltonian has the form%
\begin{eqnarray}
M&=&-t\sum_{\left\langle i,j\right\rangle ,\sigma }\left( a_{\sigma ,i}^{\dag
}b_{\sigma ,j}+h.c.\right)\nonumber\\
&-&t^{\prime }\sum_{\left\langle \left\langle
i,j\right\rangle ,\sigma \right\rangle }\left( a_{\sigma ,i}^{\dag
}a_{\sigma ,j}+b_{\sigma ,i}^{\dag }b_{\sigma ,j}+h.c.\right) ,
\label{eqn58}
\end{eqnarray}%
describing electrons that can hop from one site to the other in the
interpenetrating triangular lattices A and B \cite{castro}. The problem of
transverse vibrations of a honeycomb lattice is thus formally identical to
that of the band structure of the lattice, except for the replacement of
boson operators by fermion operators. The only difference is that while for
bosons one can put any number at each site, for fermions one can put only
one at each site (or two if one takes into accout the spin).

The full algebraic Hamiltonian for vibrations of the honeycomb lattice is%
\begin{eqnarray}
H &=&E_{0}+A\sum_{i}C_{i}+A^{\prime }\sum_{i\neq j}C_{i,j}^{\prime }+\lambda
^{(I)}\sum_{\left\langle i,j\right\rangle }M_{i,j}^{(I)} \nonumber \\
&+&\lambda^{(II)}\sum_{\left\langle \left\langle i,j\right\rangle \right\rangle
}M_{ij}^{(II)}
+\lambda ^{(III)}\sum_{\left\langle \left\langle \left\langle
i,j\right\rangle \right\rangle \right\rangle }M_{ij}^{(III)}.
\label{eqn59}
\end{eqnarray}%
There are therefore for this problem three symmetry adapter operators %
\begin{eqnarray}
S^{(I)}&=&\sum_{\left\langle i,j\right\rangle }M_{ij}^{(I)},\, 
S^{(II)}=\sum_{\left\langle \left\langle i,j\right\rangle\right\rangle
}M_{ij}^{(II)},\nonumber\\ 
S^{(III)}&=&\sum_{\left\langle \left\langle\left\langle
i,j\right\rangle \right\rangle \right\rangle }M_{ij}^{(III)}.\nonumber
\end{eqnarray}

\subsection{Solutions\label{HexagonSolutions}}

\subsubsection{The fundamental vibration v=1\label{Hexagonv1}}

The eigenvalues of the symmetry adapter operators $S^{(I)}$, $S^{(II)}$ and $%
S^{(III)}$ for the unit cell are given by%
\begin{eqnarray}
(I) &:&\, \, \pm 2,\pm 1,\pm 1  \nonumber \\
(II) &:&\, \, \pm 2,\pm 1,\pm 1  \nonumber \\
(III) &:&\, \, \pm 1,\pm 1,\pm 1,
\label{eqn60}
\end{eqnarray}%
while for the supercell they are given by%
\begin{eqnarray}
(I) &:&\,\pm 2.3810,\pm 1.7320,\pm 1.5735,\pm 1,\pm 1,\pm 0.9246,0 
\nonumber \\
(II) &:&\, 3.6457,3.0861,1,1,0.618,0.428,  \nonumber \\
&&-1,-1,-1,-1.514,-1.618,-1.645,-2  \nonumber \\
(III) &:&\,\pm \sqrt{2},\pm \sqrt{2},\pm 1,\pm 1,0,0.0,0.0.
\label{eqn61}
\end{eqnarray}
\begin{figure}[h!]
{\includegraphics[width=\linewidth]{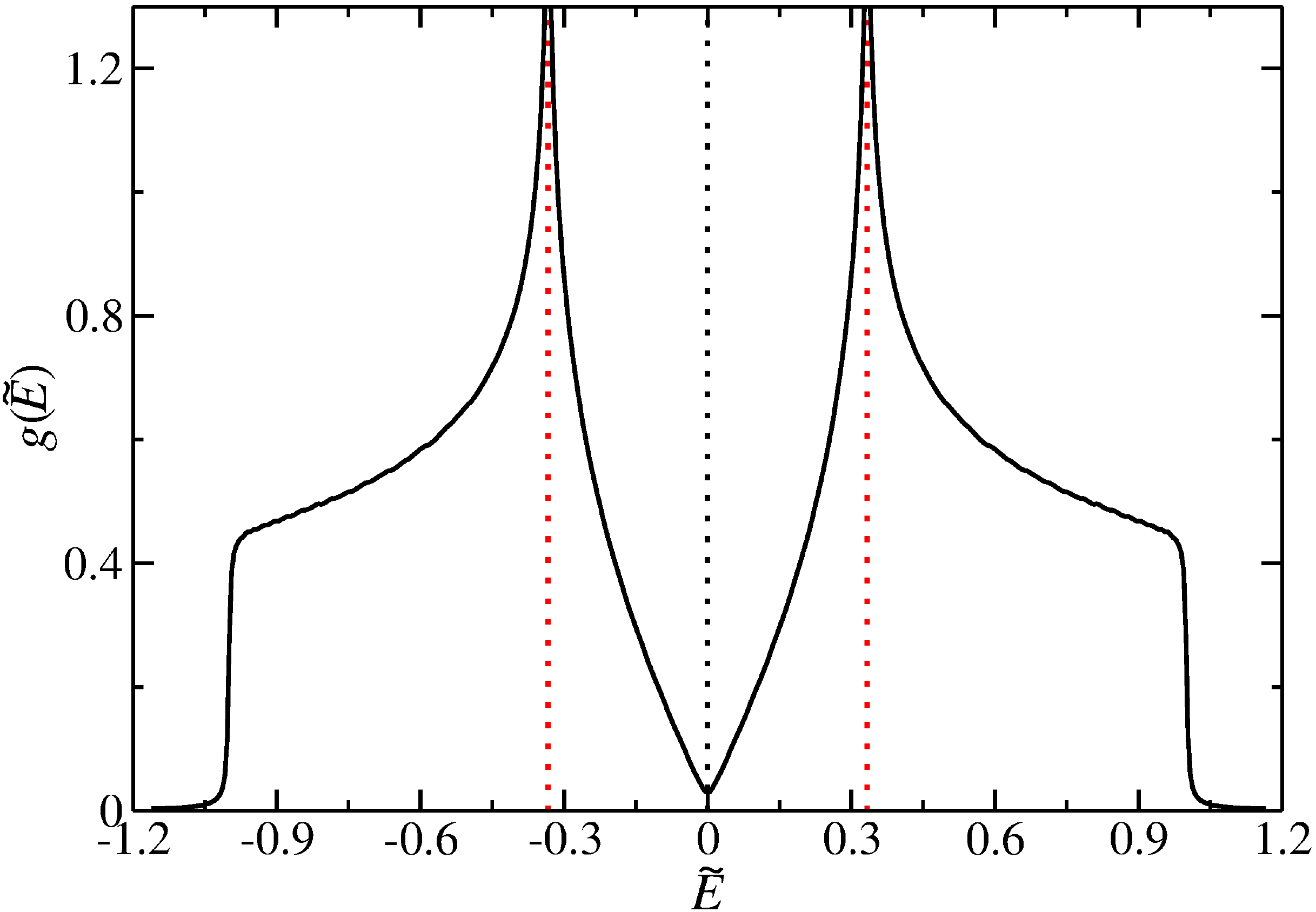}}
\caption{The DOS of the $v=1$ vibration of the honeycomb lattice, Eq.~(\ref{eqn63}).}
\label{fig14}
\end{figure}
The eigenstates are representations of $D_{6h}$~\cite{Wilson1955}. When the number of cells is
very large, an analytic solution was given by Wallace~\cite{wallace} who
solved the equivalent problem of fermions on the honeycomb lattice (see also~\cite{Reich2002}),%
\begin{eqnarray}
E\left(k_x,k_y\right)&=&\gamma_0E^{(I)}\left( k_{x},k_{y}\right)+\gamma_1E^{(II)}\left( k_{x},k_{y}\right)\\
E^{(I)}\left( k_{x},k_{y}\right) &=&\pm \sqrt{3+u(k_x,k_y)}  \nonumber \\
E^{(II)}\left( k_{x},k_{y}\right) &=&-u(k_x,k_y) \nonumber\\
u(k_x,k_y)&=&2\cos 2\pi k_{y}a+4\cos\pi k_{y}a\cos\pi k_{x}\sqrt{3}a,\nonumber
\label{eqn62}
\end{eqnarray}%
where $\pi k_{x}a=x$ and $\pi k_{y}a=y$ are continuous variables. The Hamiltonian
used by Wallace differs from $M$ by a constant (a shift in $E$). Hobson and
Nierenberg \cite{hobson} using a method similar to that described in~\refsec{Squarev1} for the square lattice provided an analytic expression for
the DOS for nearest-neighbor interactions. Introducing $%
E_{\max }=6\lambda ^{(I)}$ and $\tilde{E}=E/E_{\max },$ $-1<\tilde{E}<1$, $\omega =(\tilde E+1)/2$, the DOS is given in terms of an elliptic integral%
\begin{eqnarray}
g(\omega)=&\frac{9}{\pi^{2}}\sqrt{\frac{1-2\omega}{3}}K\left(\sqrt{\frac{\omega(2-3\omega)^{3}}{(1-2\omega)}}\right),\, &0<\omega<\frac{1}{3},  \nonumber \\
g(\omega)=&\frac{9}{\pi ^{2}}\frac{(1-2\omega)}{\sqrt{3\omega(2-3\omega)^{3}}}K\left(\sqrt{\frac{1-2\omega}{\omega(2-3\omega)^3}}\right),\, &\frac{1}{3}\leq\omega\leq\frac{1}{2},
\label{eqn63}
\end{eqnarray}%
and the function is symmetric around $\omega=\frac{1}{2}$, i.e., around $\tilde E=0$, see Fig.~\ref{fig14}.

\begin{figure}[h!]
{\includegraphics[width=0.8\linewidth]{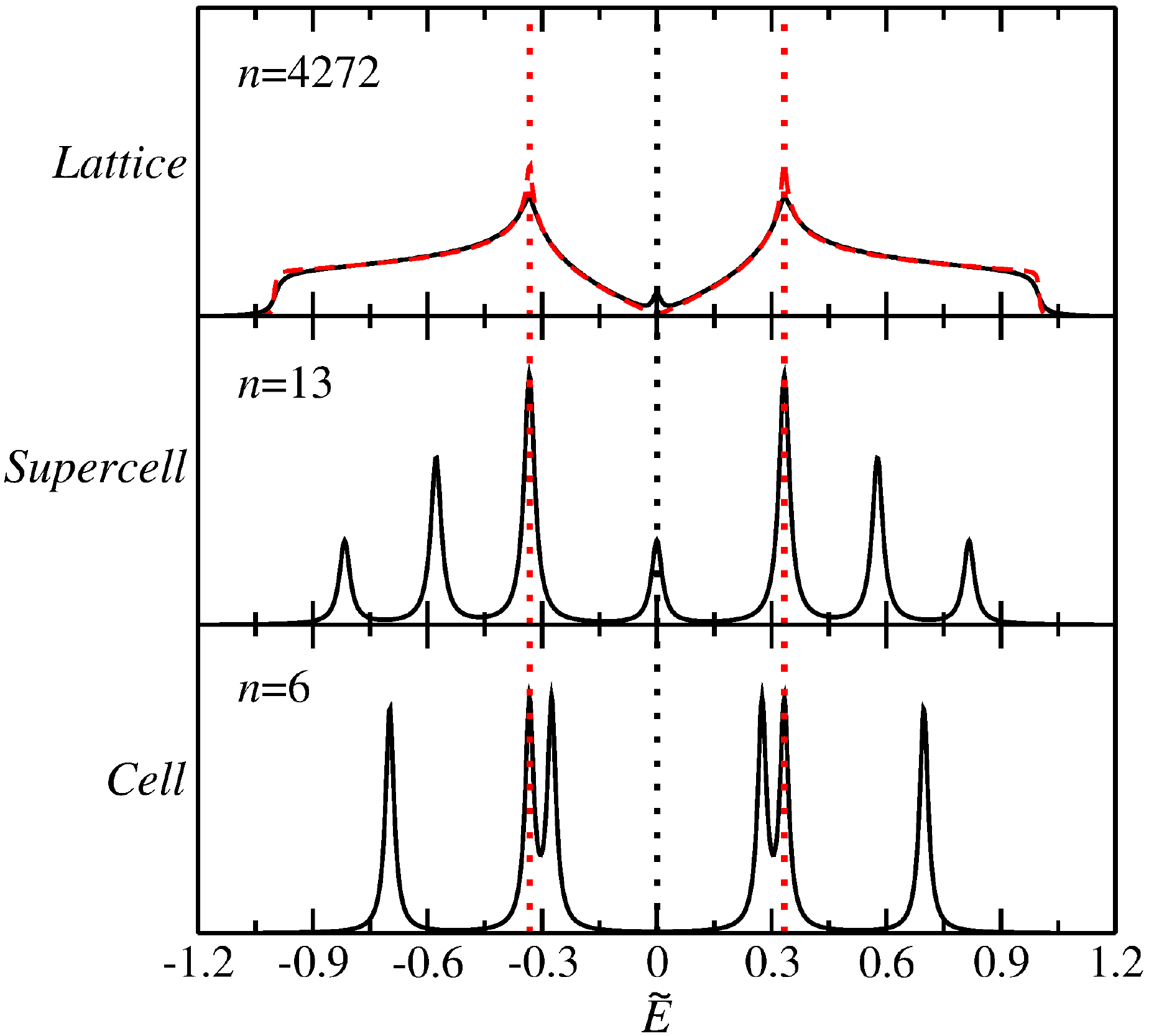}}
\caption{DOS of the fundamental vibration $v=1$ of the honeycomb lattice including the unit cell and the supercell with the number of sites $n$ as indicated in the insets. Red dashed line: Eq.~(\ref{eqn63}).}
\label{fig15}
\end{figure}

The Hamiltonian used by Hobson and Nierenberg differs from $M$, and from the
Hamiltonian of Wallace by a constant. The DOS Eq.~(\ref{eqn63}) should be compared with
that of a square lattice, given by Eq.~(\ref{eqn48}). The difference between the two is a
consequence of the symmetry of the lattice. The major difference between the
two DOSs is the occurrence of a zero (Dirac zero) at $\tilde{E}=0$ in the honeycomb lattice. The properties of the DOS of the honeycomb
lattice are due to the symmetry $D_{6h}$ of the unit cell and are already
encoded into the DOS of the unit cell and the supercell, as shown in Fig.~\ref{fig15}.

\begin{figure}[h!]
{\includegraphics[width=0.8\linewidth]{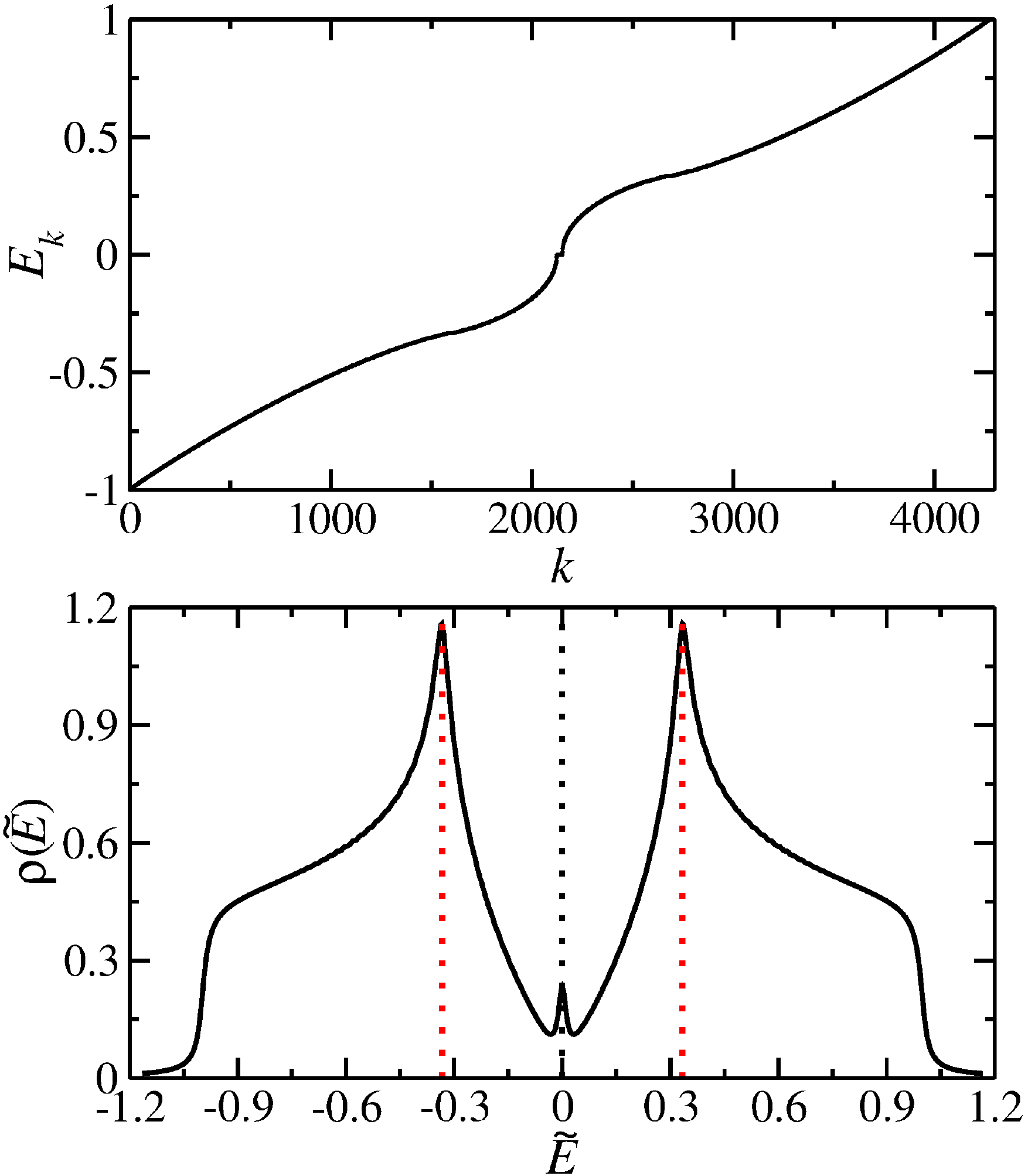}}
\caption{EDR and DOS of the fundamental vibration $v=1$ of a honeycomb lattice with $n=4272$ sites.}
\label{fig16}
\end{figure}

These results have been verified by numerical simulations as shown in Fig.~\ref{fig16}. For finite $n$ a peak occurs at $E=0$, which is due to the presence of zigzag edges in the lattice structure~\cite{Wurm2011}.

\subsubsection{The overtone v=2\label{Hexagonv2}}

\begin{figure}[h!]
{\includegraphics[width=0.8\linewidth]{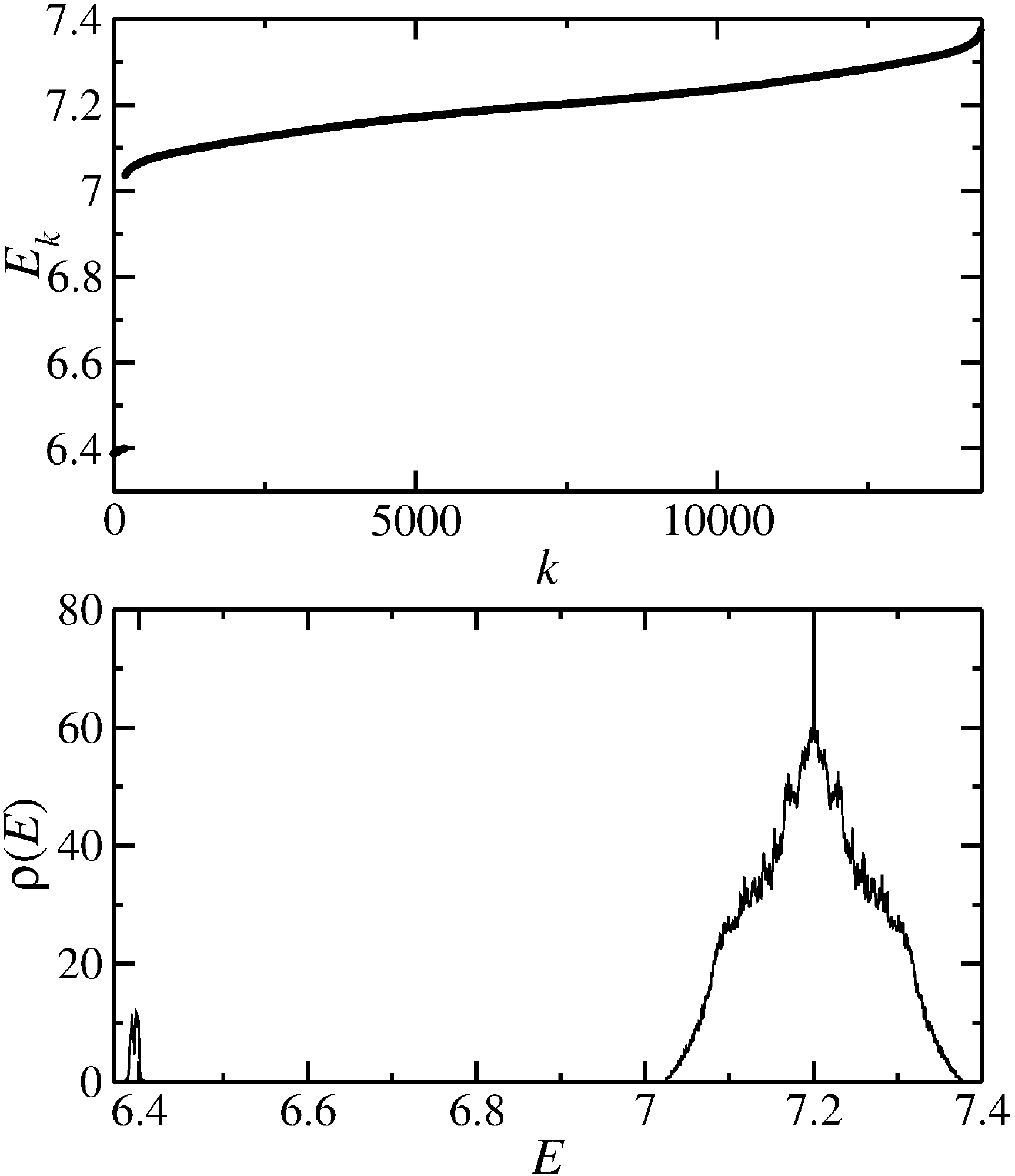}}
\caption{EDR and DOS of the overtone vibration $v=2$ of a honeycomb lattice with $n=169,\, A=-1,\, N=10,\, \lambda=0.03$.}
\label{fig17}
\end{figure}

\begin{figure}[h!]
{\includegraphics[width=\linewidth]{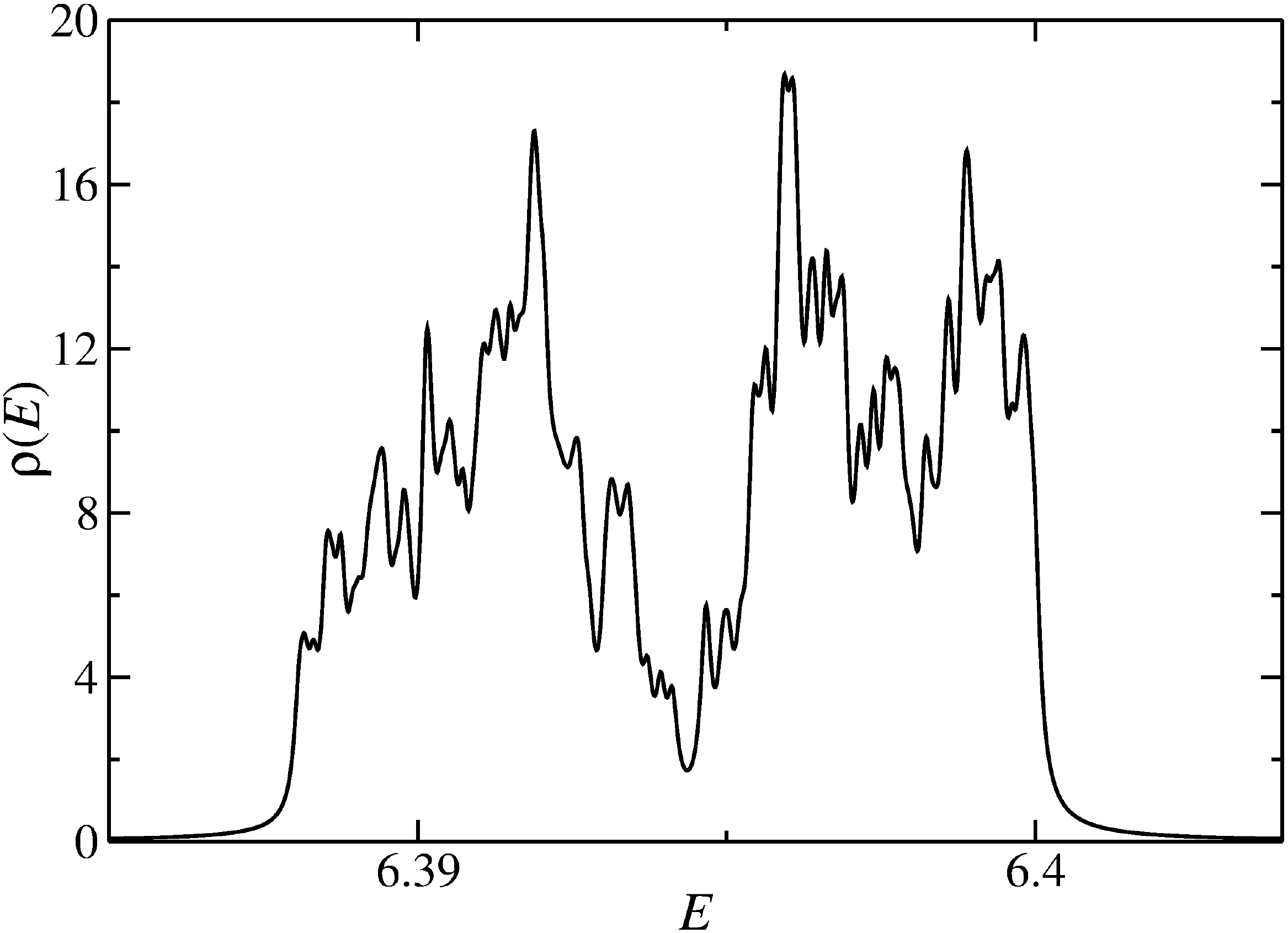}}
\caption{Zoom into the small peak in the DOS of the overtone vibration $v=2$ of a honeycomb lattice with $n=169,\, A=-1,\, N=10,\, \lambda=0.03$ shown in Fig.~\ref{fig17}.}
\label{fig17a}
\end{figure}

\begin{figure}[h!]
{\includegraphics[width=0.8\linewidth]{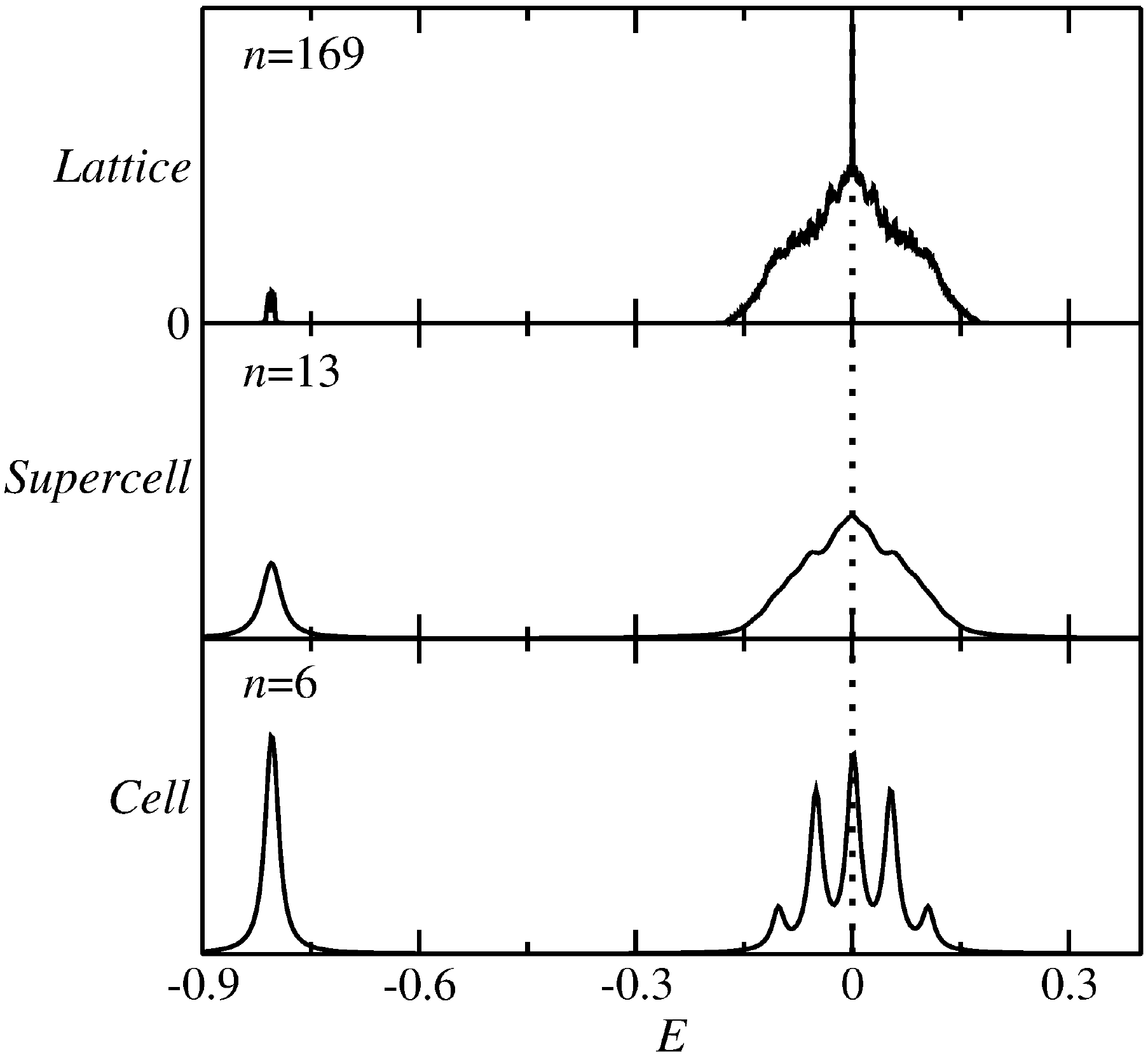}}
\caption{DOS of the $v=2$ overtone of a honeycomb lattice including the cell and supercell with the number of sites $n$ indicated in the insets and $A=-1,\, N=10,\, \lambda=0.03$.}
\label{fig18}
\end{figure}

The spectrum (EDR) and density of states (DOS) can also be calculated
by diagonalizing the Hamiltonian $H$ of Eq.~(\ref{eqn56}) in the space $v=2$, $%
\left\vert 2,0,0,...\right\rangle $, $\left\vert 1,1,0,...\right\rangle $, $%
\left\vert 1,0,1,0,...\right\rangle ,$ $...$., in terms of the five
parameters $A$, $A^{\prime }$, $\lambda ^{(I)}$, $\lambda ^{(II)}$, $\lambda
^{(III)}$, and the anharmonicity parameter, $N$. In order to illustrate the
properties of the solutions we consider the case of a highly anharmonic
vibration with $N=10$, $-4A=4$, $-4A^{\prime }=0$, and only nearest-neighbor
interactions, $\lambda ^{(I)}=0.03$, $\lambda ^{(II)}=\lambda ^{(III)}=0$.
Also here the DOS of $v=2$ has two parts, as shown in Fig.~\ref{fig17}. The first
part diverges as $n$ and has only edge singularities as illustrated in Fig.~\ref{fig17a}, the second part
diverges as $n^{2}$, and has a logarithmic singularity and two shoulders.
The splitting into these two parts can also be seen in the unit cell and in
the supercell as shown in Fig.~\ref{fig18}.

\section{Summary of DOS and EDR\label{Summary}}

\begin{figure}[h!]
{\includegraphics[width=\linewidth]{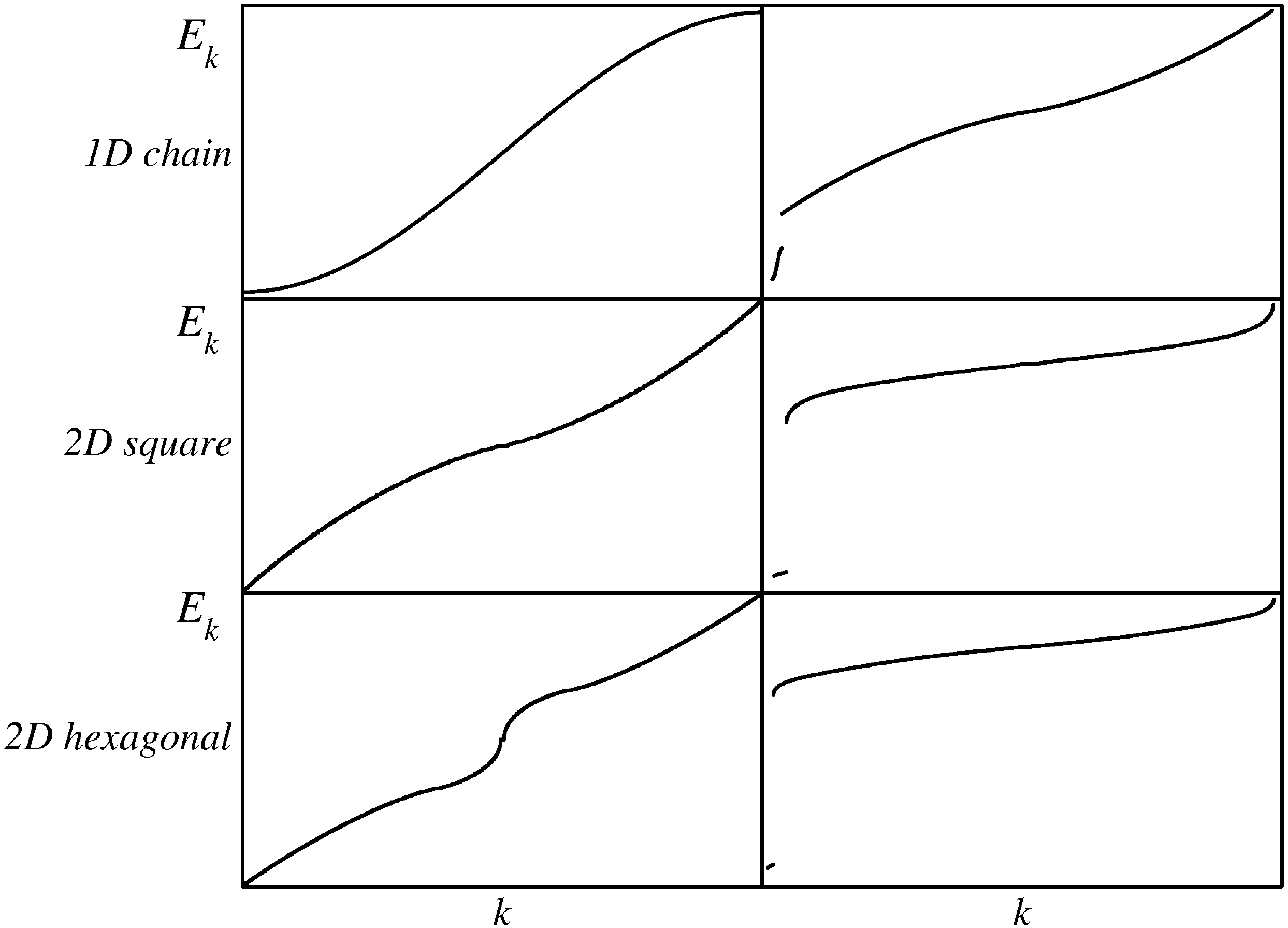}}
\caption{Summary of the EDRs of transverse crystal vibrations. Left: $v=1$, right: $v=2$.}
\label{fig19}
\end{figure}

\begin{figure}[h!]
{\includegraphics[width=\linewidth]{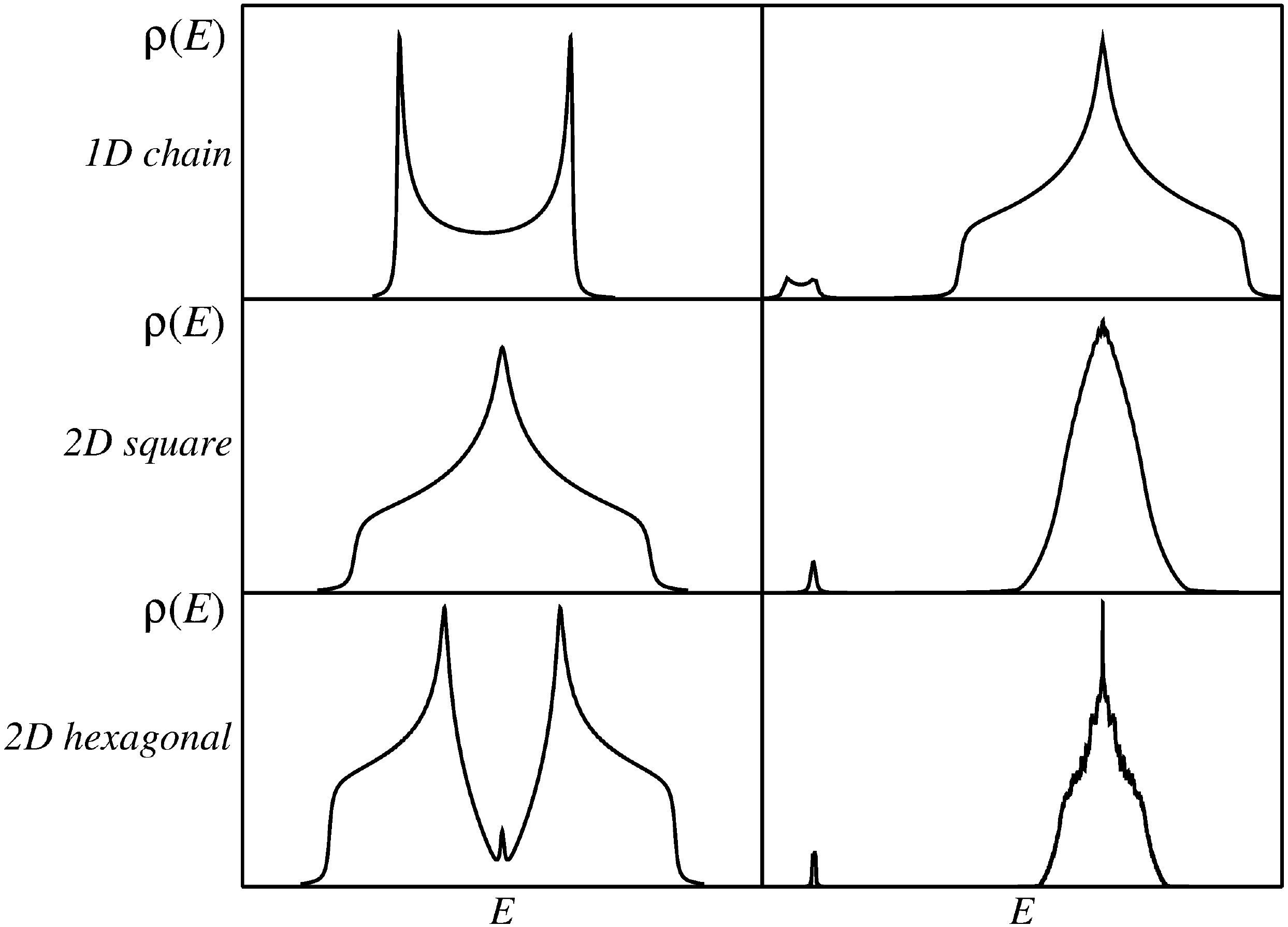}}
\caption{Same as Fig.~\ref{fig19} for the DOEs.}
\label{fig20}
\end{figure}

The results of sections~\ref{Chain},~\ref{Square} and~\ref{Hexagon} are summarized in Figs.~\ref{fig19} and~\ref{fig20}. For the
fundamental vibration $v=1$, they confirm the conjecture of van Hove \cite%
{vanhove} that singularities in the level density occur only in two
dimensions.
\section{Energy surfaces and phonon dispersion relations}

\begin{figure}[h!]
{\includegraphics[width=0.5\linewidth]{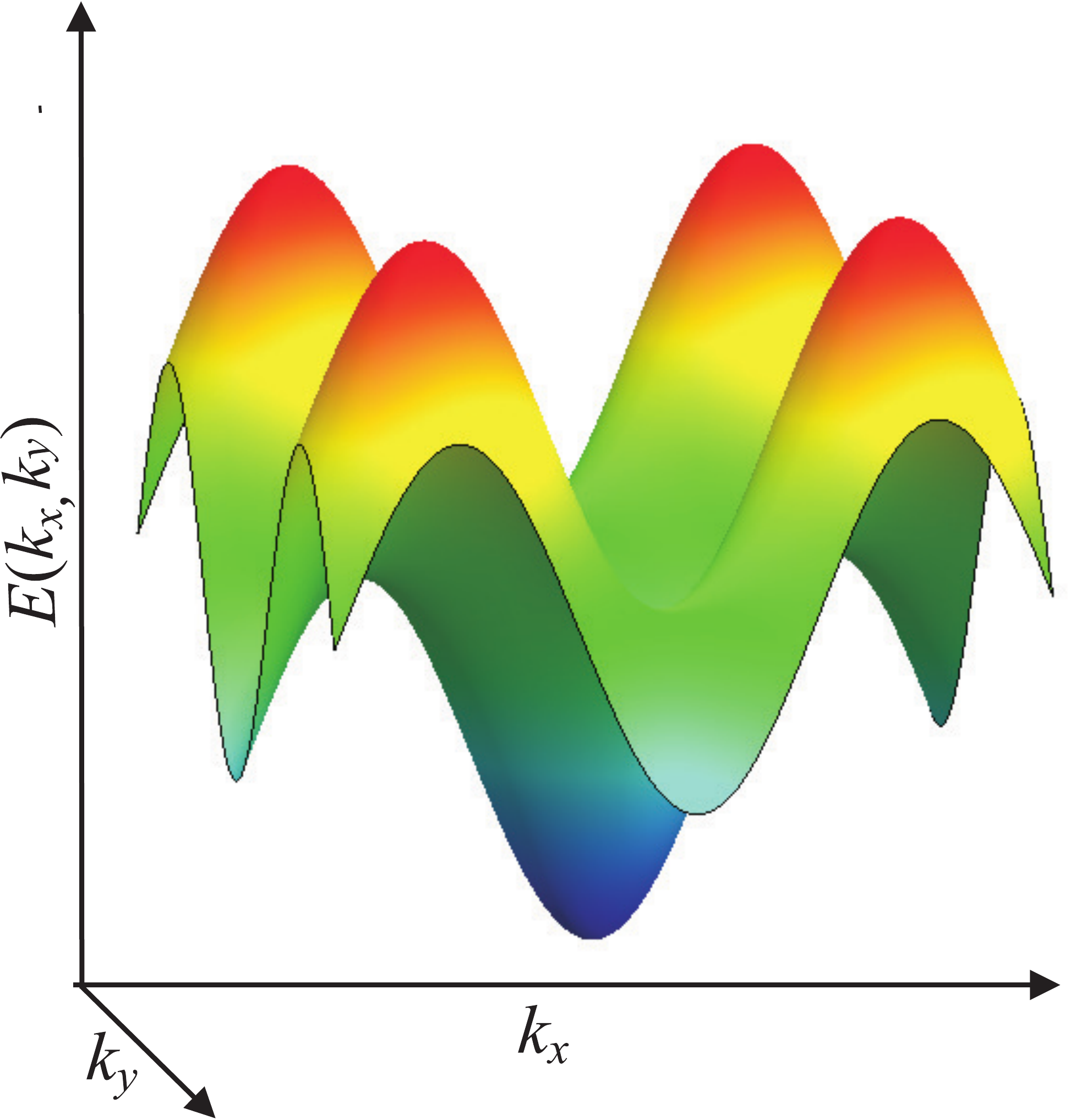}}
{\includegraphics[width=0.4\linewidth]{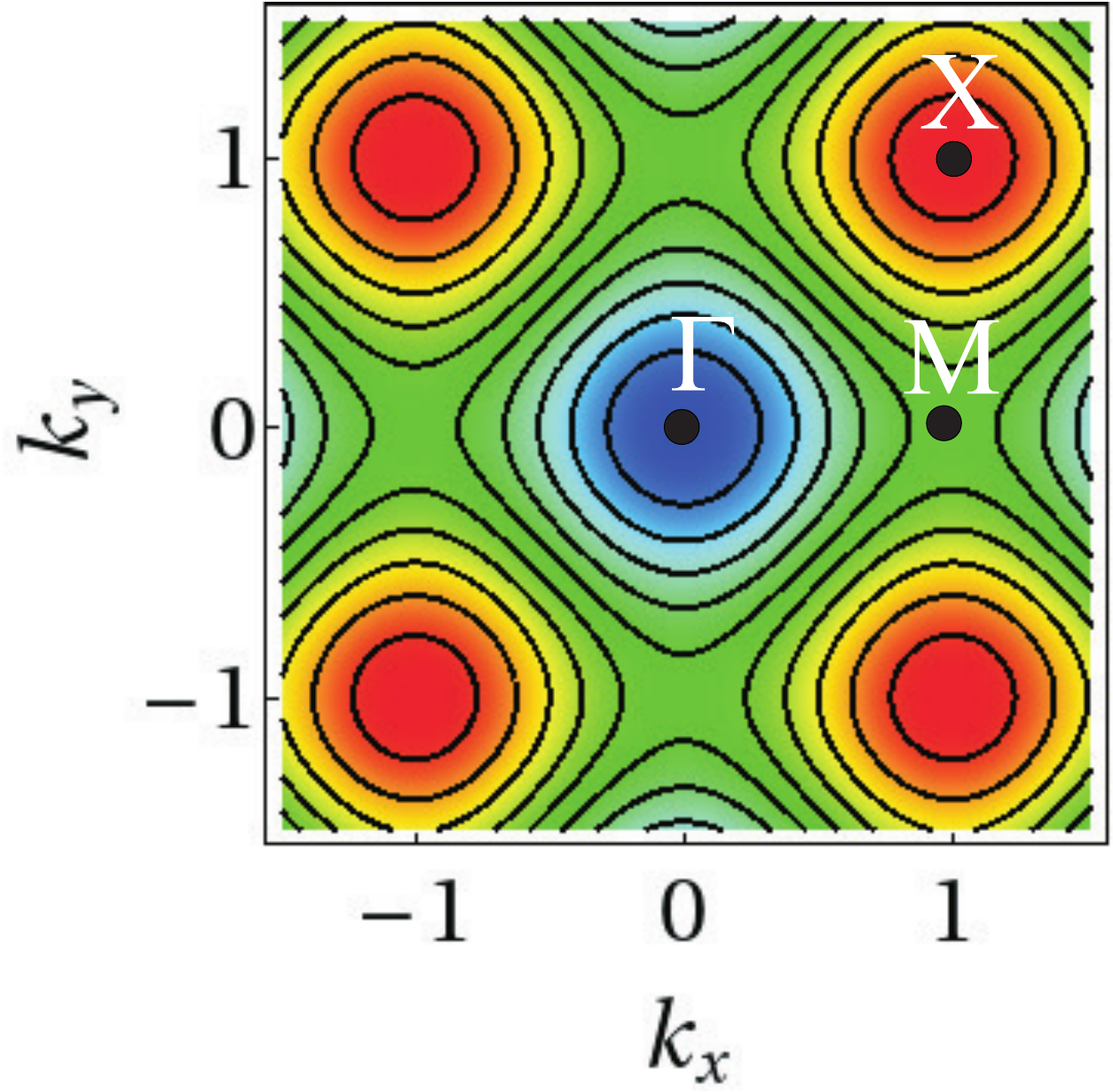}}
\caption{The left panel shows the energy surface of the $v=1$ vibrations, Eq.~(\ref{ESquare}) of an infinite-size square lattice, where the lattice constant $a$ was set to unity. The right panel shows the corresponding density plot in the quasimomentum plane $(k_x,k_y)$ with the isofrequency lines shown as dark lines.}
\label{fig24}
\end{figure}

\begin{figure}[h!]
{\includegraphics[width=\linewidth]{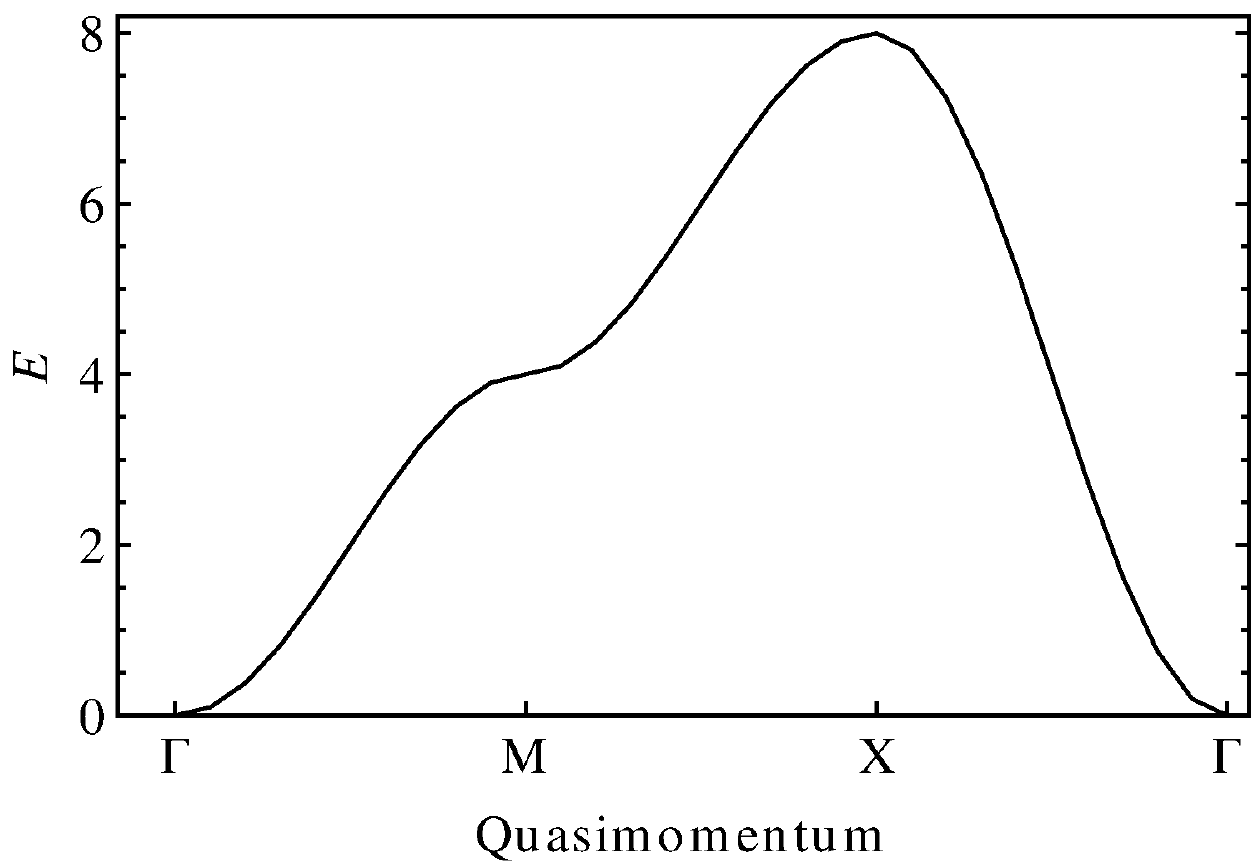}}
\caption{Computed phonon dispersion relation of an infinite-size square lattice.}
\label{fig25}
\end{figure}

It is of interest to display explicitly the energy surfaces and phonon dispersion
relations of infinite-size lattices. For the square lattice, the energy surface for 
nearest-neighbor interactions is given by Eq.~(\ref{eqn45}), rewritten, for $2\lambda^{(I)}=1$, as
\begin{equation}
E(k_x,k_y) = (2 - \cos\pi k_xa-\cos\pi k_ya).
\label{ESquare}
\end{equation}
This energy surface is shown in Fig.~\ref{fig24}.
The corresponding phonon dispersion relation along the boundary of the first irreducible Brillouin zone is shown in Fig.~\ref{fig25}. 
The special points $\Gamma(0,0)$, ${\rm M}\left(\frac{1}{2a},0\right)$, ${\rm X}\left(\frac{1}{2a},\frac{1}{2a}\right)$ are indicated in the density plot, shown in the right part of Fig.~\ref{fig24}.

\begin{figure}[h!]
{\includegraphics[width=0.5\linewidth]{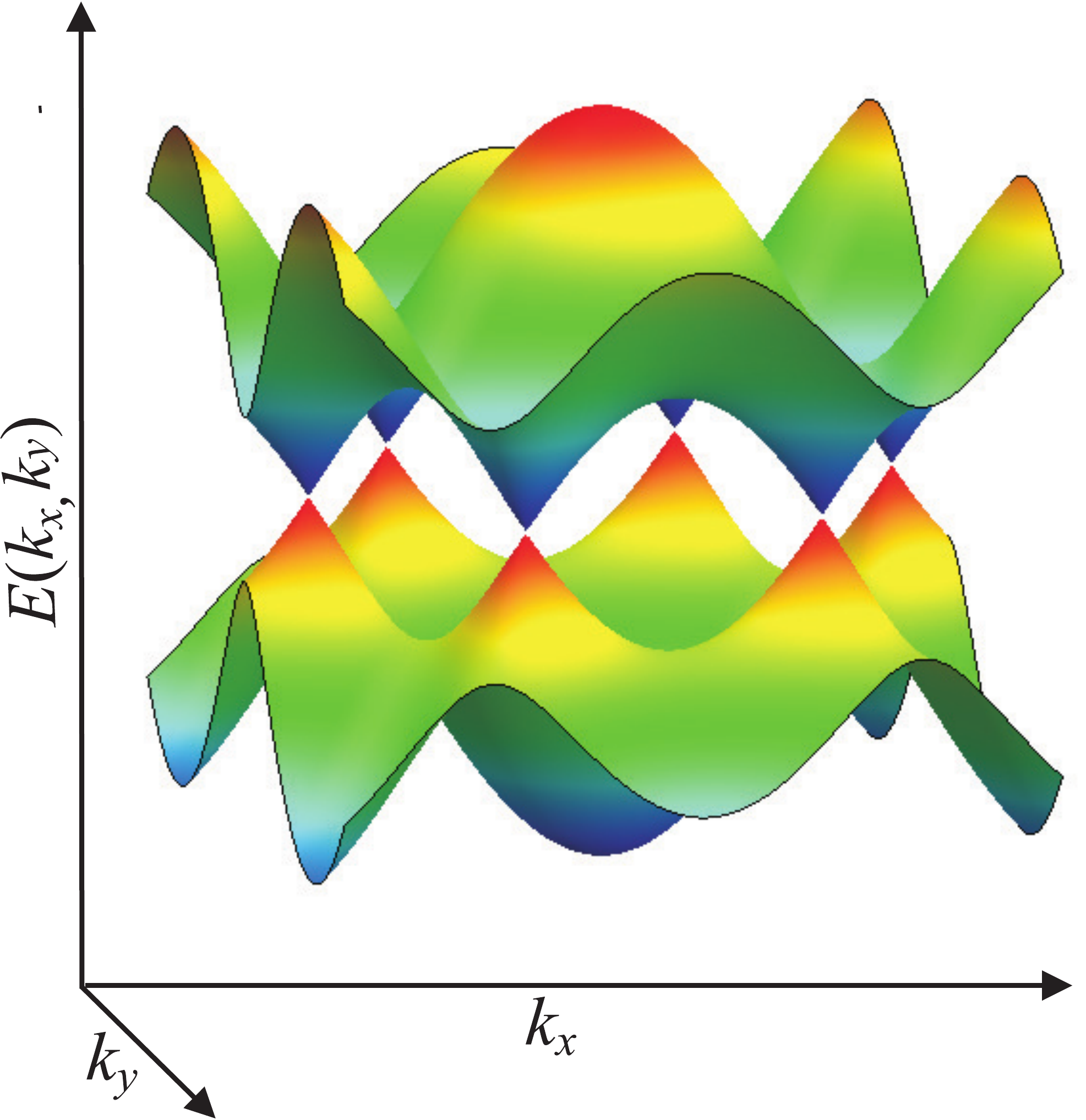}}
{\includegraphics[width=0.4\linewidth]{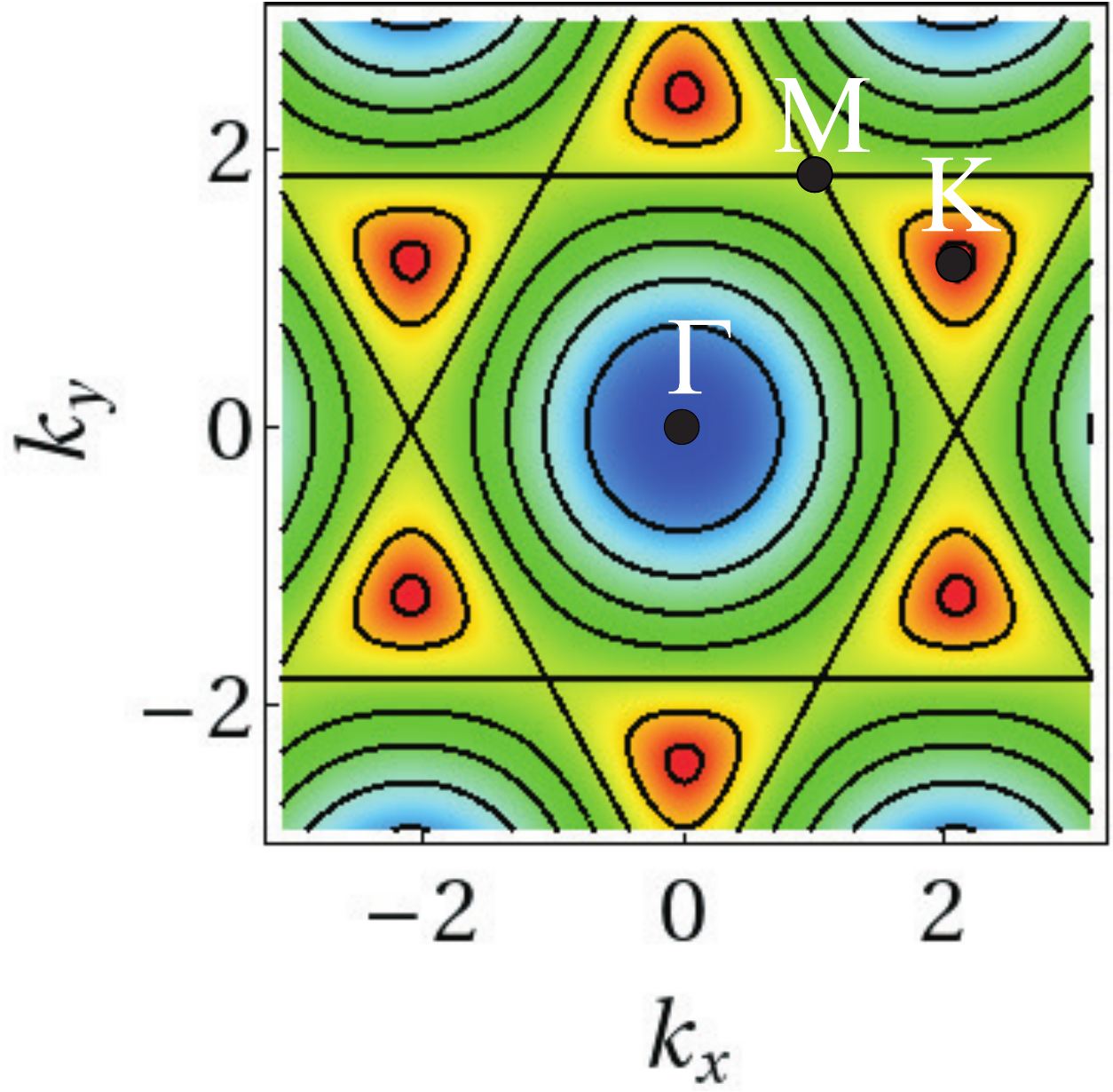}}
\caption{The left panel shows the two sheets of the energy surface of the $v=1$ vibrations, Eq.~(\ref{EHexagon}) of an infinite-size hexagonal lattice, where $a$ was set to unity. They touch each other conically at the corners of the first Brillouin zone. The right panel shows the corresponding density plot in the quasi-momentum plane $(k_x,k_y)$ with the isofrequency lines shown as dark lines.}
\label{fig26}
\end{figure}

\begin{figure}[h!]
{\includegraphics[width=\linewidth]{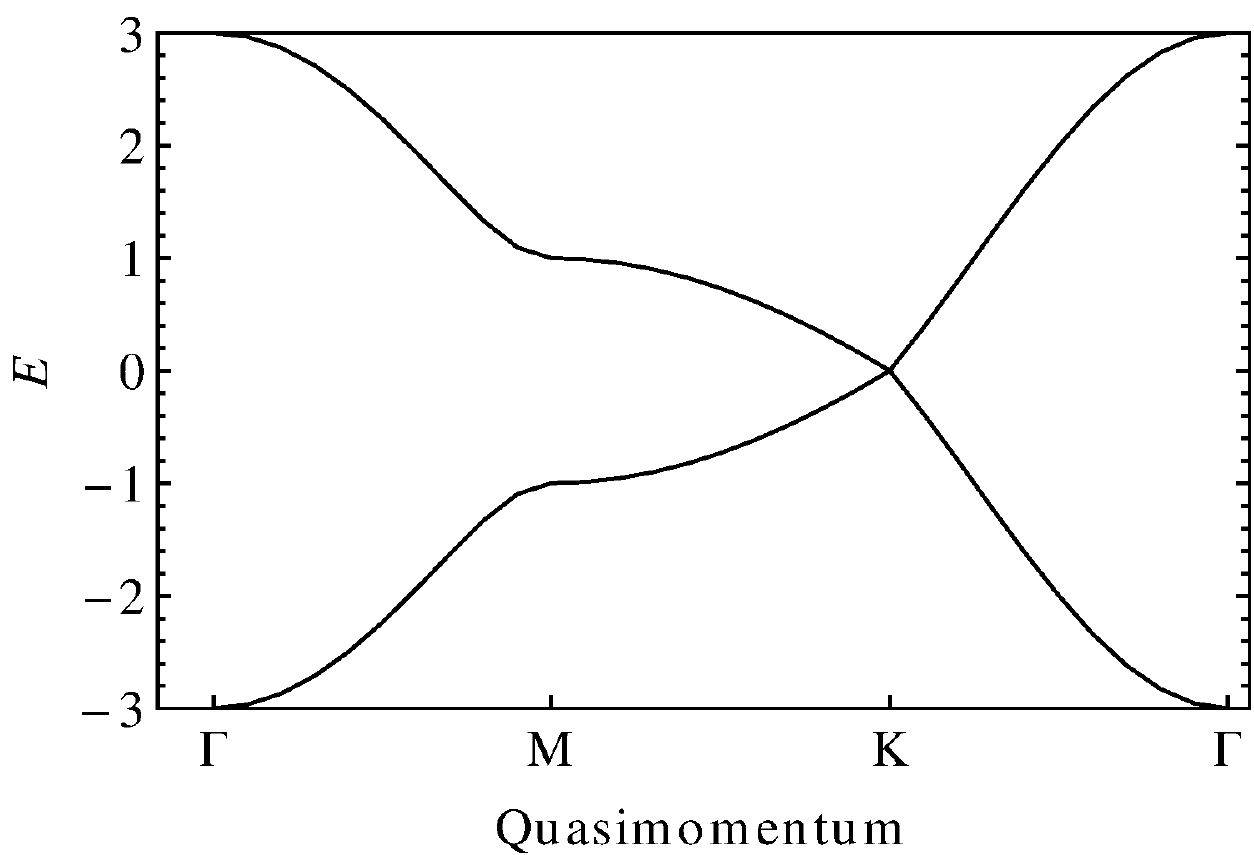}}
\caption{Computed phonon dispersion relation of an infinite-size hexagonal lattice.}
\label{fig27}
\end{figure}

For the hexagonal lattice, the energy surface for nearest-neighbor interaction
is given by Eq.(62) rewritten as~\cite{castro}
\begin{eqnarray}
E (k_x,k_y)&=&\pm\sqrt{3 + f(k_x,k_y)}
\label{EHexagon}\\
f(k_x,k_y)&=&2\cos\left(\sqrt{3}k_ya\right) + 4\cos\left(\frac{\sqrt{3}}{2}k_ya\right)
\cos\left(\frac{3}{2}k_xa\right).\nonumber
\end{eqnarray}
This energy surface is shown in Fig.~\ref{fig26}. 
Because of the $\pm$ sign in Eq.(\ref{EHexagon}), it consists of two sheets which touch conically at the 
corners of the first Brillouin zone. This $\pm$ sign is a consequence of the fact that the hexagonal lattice 
can be viewed as two interpenetrating triangular lattices as discussed in \refsec{Hexagon}. This
is the crucial property that makes the hexagonal lattice so different from other lattices. The phonon 
dispersion relation is shown in Fig.~\ref{fig27}. 
The special points $\Gamma(0,0)$, ${\rm M}\left(\frac{2\pi}{3a},0\right)$, ${\rm K}\left(\frac{2\pi}{3a},\frac{2\pi}{3\sqrt{3}a}\right)$ are indicated in the density plot, shown in the right part of Fig.~\ref{fig26}. In Figs.~\ref{fig24} and~\ref{fig26} one can see clearly the symmetry of the unit cells $D_{4h}$ (square) and $D_{6h}$ (hexagonal).

As mentioned in \refsec{Hexagon}, the structure of the algebraic Hamiltonian applies
to both bosons and fermions, and it depends only on the symmetry of the lattice.
Thus, the energy surfaces and dispersion relations, Eqs.~(\ref{ESquare}) and~(\ref{EHexagon}), apply also to
electrons in a square and hexagonal lattice. In this case, $\boldsymbol{k}=(k_x,\, k_y)$ represents the
quasimomentum of electrons. This is what makes graphene so special.
\section{Comparison with data in microwave photonic crystals\label{Experiment}}

Superconducting microwave resonators have been used for two decades as analog
systems for the study of quantum phenomena in high resolution measurements 
\cite{billiards,crystals,crystals2,Chaos2015}. Photonic crystals are the optical analog
of a solid. Both concepts can be combined into "microwave photonic crystals"
which offer the opportunity to perform high precision measurements of the
excitation spectrum, and thus to study the EDR and DOS of solids. In
particular, recently two-dimensional hexagonal structures were realized by squeezing a photonic crystal, which was composed of several hundreds of metallic cylinders forming a triangular lattice, between two metal plates~\cite{dietz1,dietz2,Dietz2015}. 

\begin{figure}[h!]
\includegraphics[width=0.4\linewidth,height=0.25\linewidth]{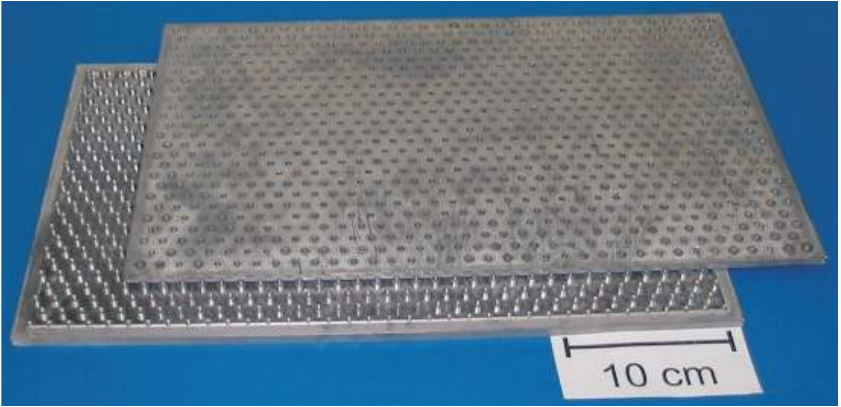}
\includegraphics[width=0.4\linewidth,height=0.25\linewidth]{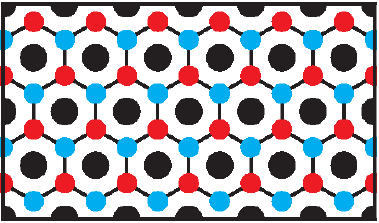}
\caption{(Color online) Left part: Photograph of the rectangular microwave photonic crystal, which contains $888$ metal cylinders arranged on a triangular grid. The top plate was shifted with respect to the bottom one for presentational reasons. Right part: Schematic view of the triangular lattice. The red (gray) and blue (dark gray) dots mark the voids between the cylinders that form the hexagonal lattice. Adopted from~\cite{Chaos2015}}
\label{photosdiracrect}
\end{figure}

\begin{figure}[h!]
\includegraphics[width=0.7\linewidth]{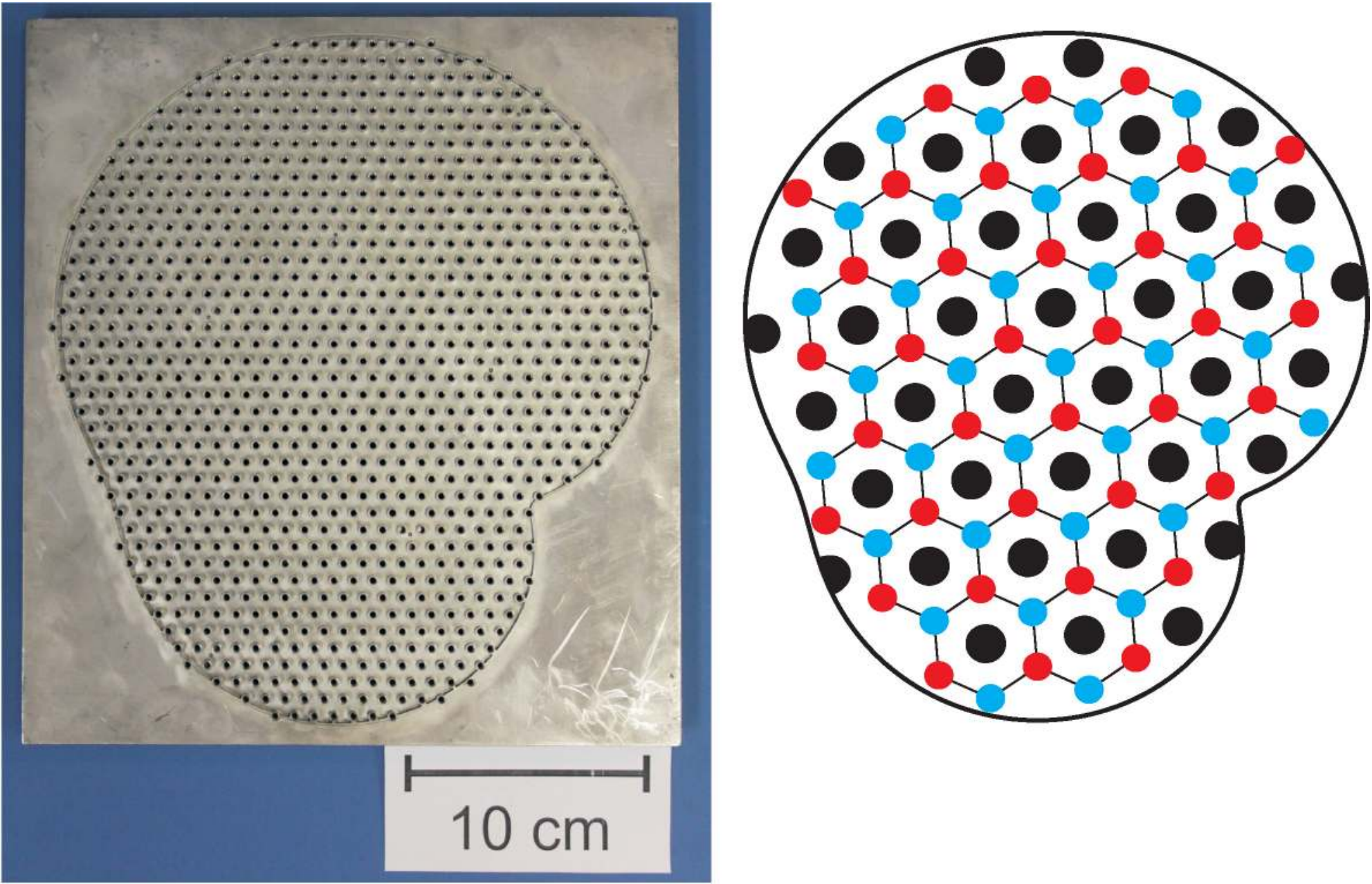}
\caption{(Color online) Same as in Fig.~\ref{photosdiracrect} but for an Africa billiard. The top plate was removed. Adopted from~\cite{Chaos2015}}
\label{photosdiracafr}
\end{figure}

In order to investigate the EDR and the DOS of hexagonal lattices of varying shapes high-precision experiments were performed with flat, superconducting microwave resonators with the forms of a rectangle and the African continent~\cite{Berry1987}, respectively. Photographs are shown in the left parts of Figs.~\ref{photosdiracrect} and~\ref{photosdiracafr}. 
Both resonators consisted of a basin and a lid, that were made from brass plates and then lead plated to attain superconductivity at the liquid helium temperature of 4.2 K. Note that the critical temperature of lead is $T_c=7.2~K$. For the construction of the photonic crystals located inside the basins $\approx 900$ cylinders were milled out of the bottom plate. The cylinders were arranged on a triangular grid, as indicated in the schematic views in Figs.~\ref{photosdiracrect} and~\ref{photosdiracafr}. Then the voids at the centers of the cells formed by, respectively, three of them yield a hexagonal configuration. The height of the resonators was $d=3$~mm and the range of excitation frequences $f$ of the microwaves that were coupled into the resonator was chosen as $0\leq f\leq f_{max}=c/2d$ so the electric field vector was perpendicular to the top and bottom plates of the resonators. Then the microwaves inside the resonators are governed by the scalar Helmholtz equation for the electric field strength with Dirichlet boundary conditions at the walls of the cylinders and the basin and the microwave photonic crystals correspond to experimental realizations of hexagonal lattices. Consequently, the model Eq.~(\ref{eqn58}) should be applicable. 

\begin{figure}[h!]
{\includegraphics[width=\linewidth]{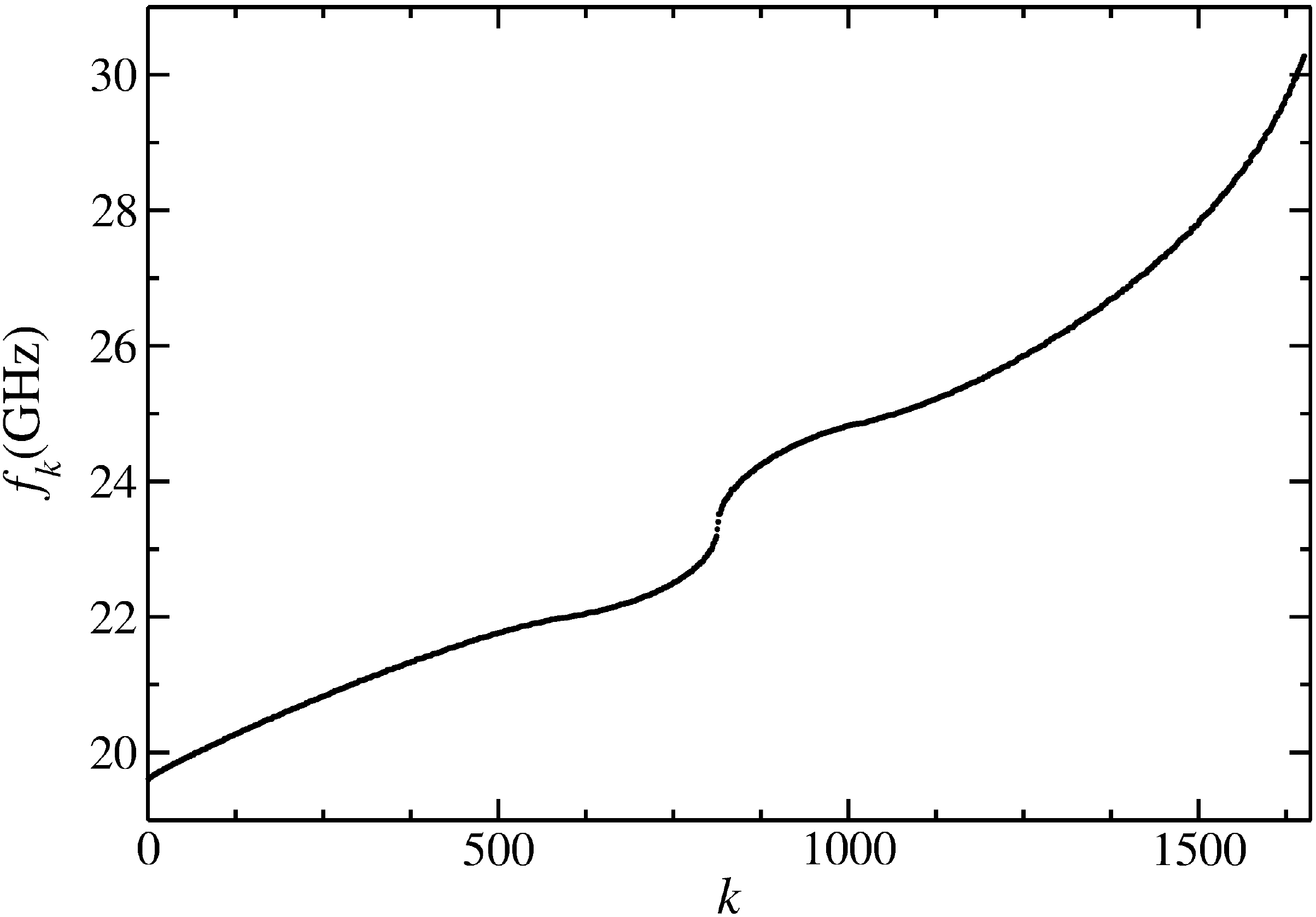}}
{\includegraphics[width=\linewidth]{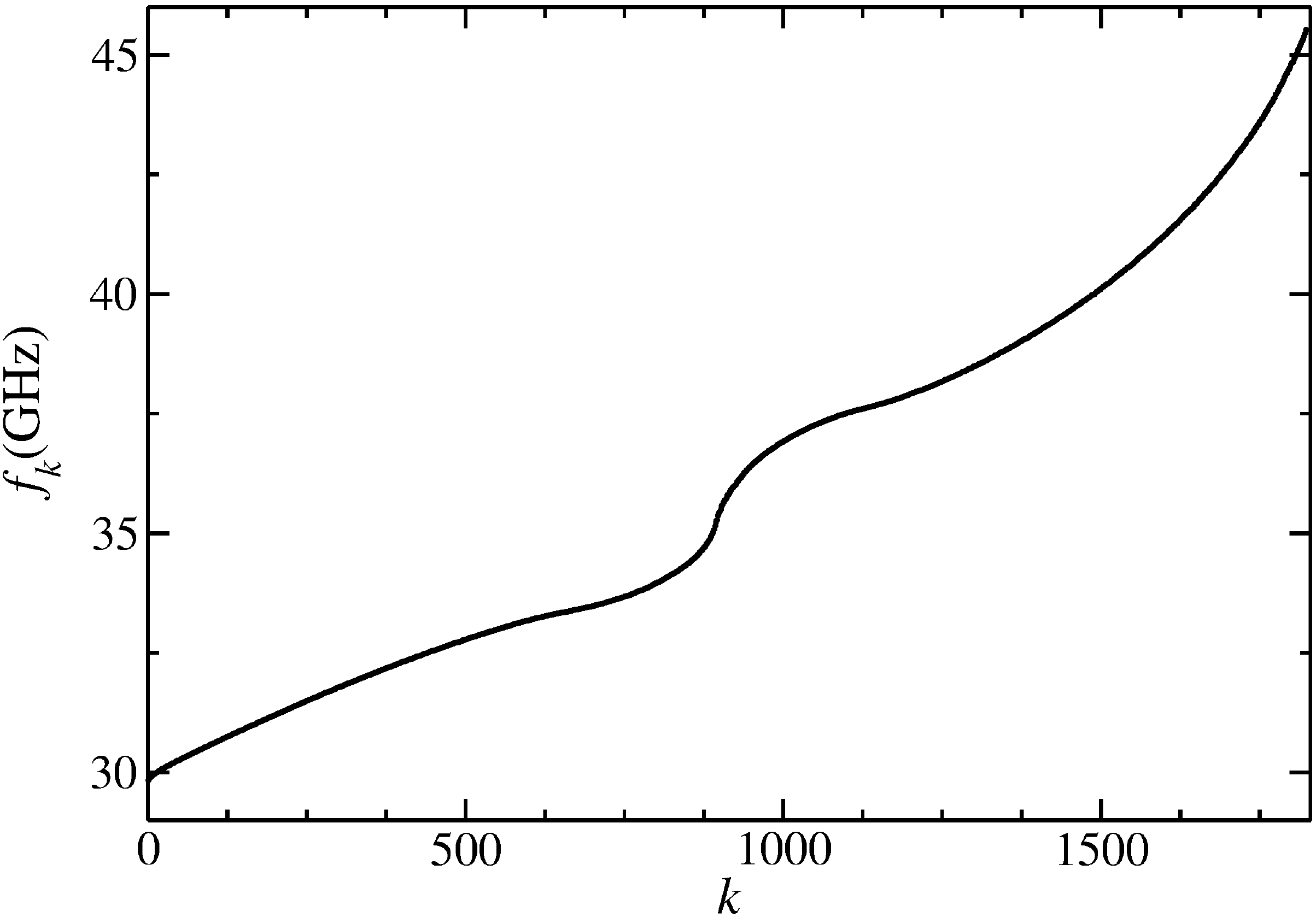}}
\caption{Experimental EDR for the rectangular photonic crystal with 1656 sites (upper panel) and the Africa-shaped one with 1823 sites (lower panel).}
\label{fig21}
\end{figure}

\begin{figure}[h!]
{\includegraphics[width=\linewidth]{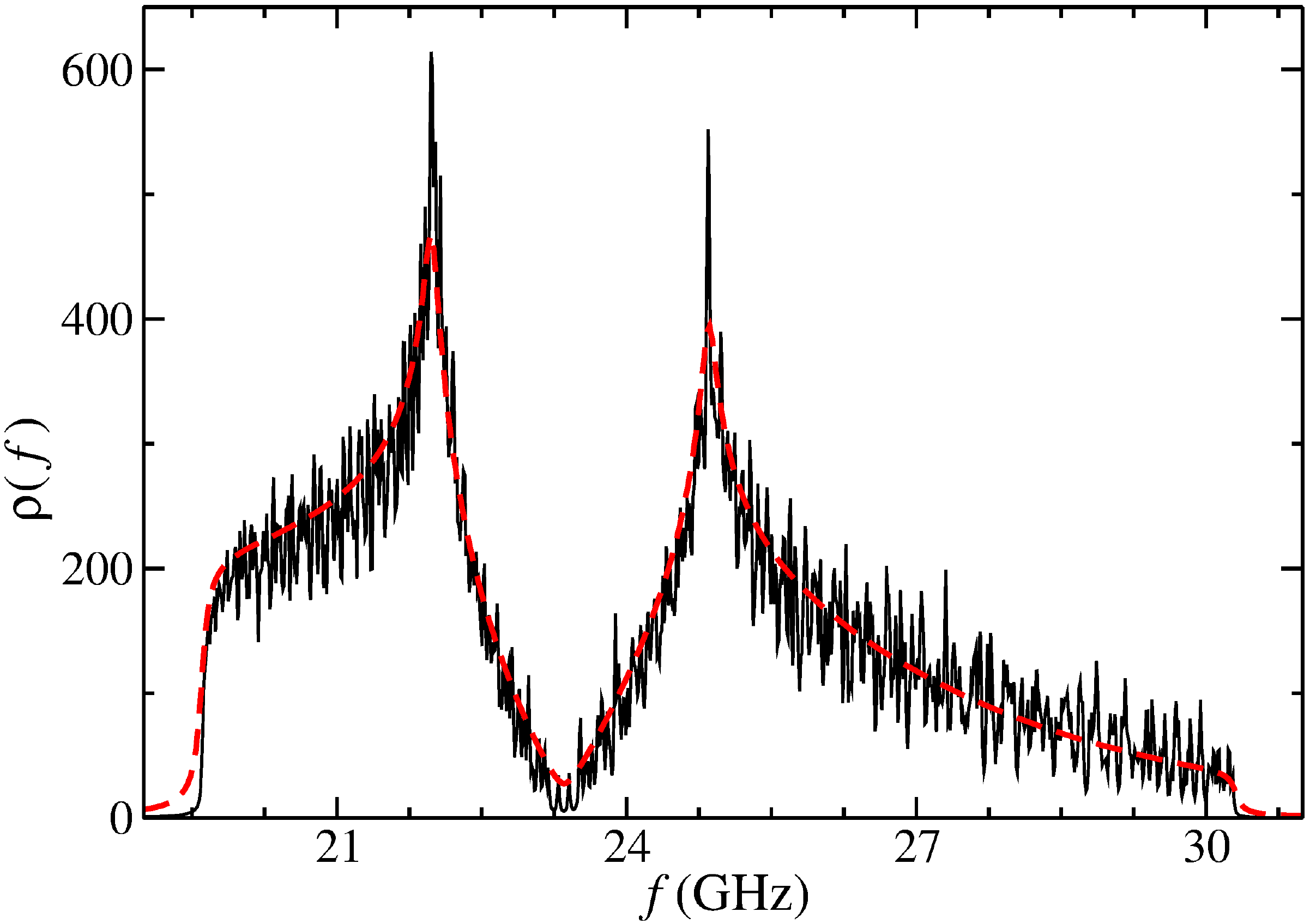}}
{\includegraphics[width=\linewidth]{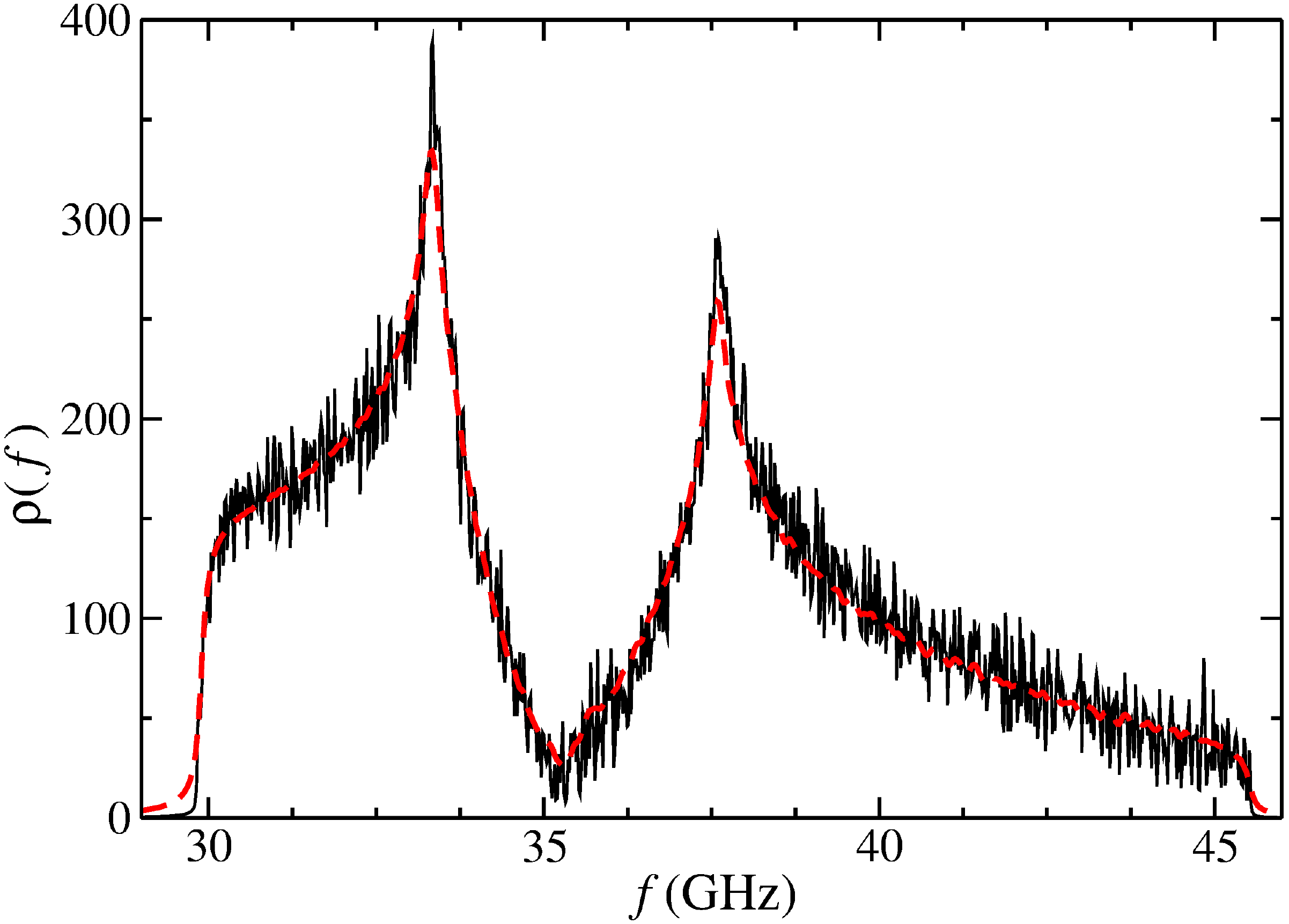}}
\caption{Same as Fig.~\ref{fig21} for the experimental DOS. The red dashed lines show the fits of the DOS deduced from the Hamiltonian in Eq.~(\ref{eqn59}) to the experimental ones.}
\label{fig22}
\end{figure}

We determined altogether 1651 and 1823 resonance frequencies in the first two bands of the rectangular and the Africa-shaped microwave photonic crystal, respectively. 
The EDRs, $f_k$ vs. $k$, and the DOSs $\rho \left( f\right)$ are shown in Figs.~\ref{fig21} and~\ref{fig22}. A fit to the DOS with the algebraic Hamiltonian Eq.~(\ref{eqn59}) with nearest-neighbor
coupling, $\lambda ^{(I)}$, \ and second- and third-neighbor couplings,
$\lambda ^{(II)}$ and $\lambda ^{(III)}$ yielded the red dashed curves in Fig.~\ref{fig22}. The best fit values are for the rectangular crystal $\lambda^{(I)}=4.573$~GHz, $\lambda^{(II)}=-0.284$~GHz, $\lambda^{(III)}=0.104$~GHz, and for the crystal with the shape of Africa $\lambda^{(I)}=2.198$~GHz, $\lambda^{(II)}=0.107$~GHz, $\lambda^{(III)}=-0.013$. The data exhibit both the van Hove singularities~\cite{vanhove} and a vanishing DOS at the Dirac frequency~\cite{crystals2}.
\begin{figure}[h!]
{\includegraphics[width=\linewidth]{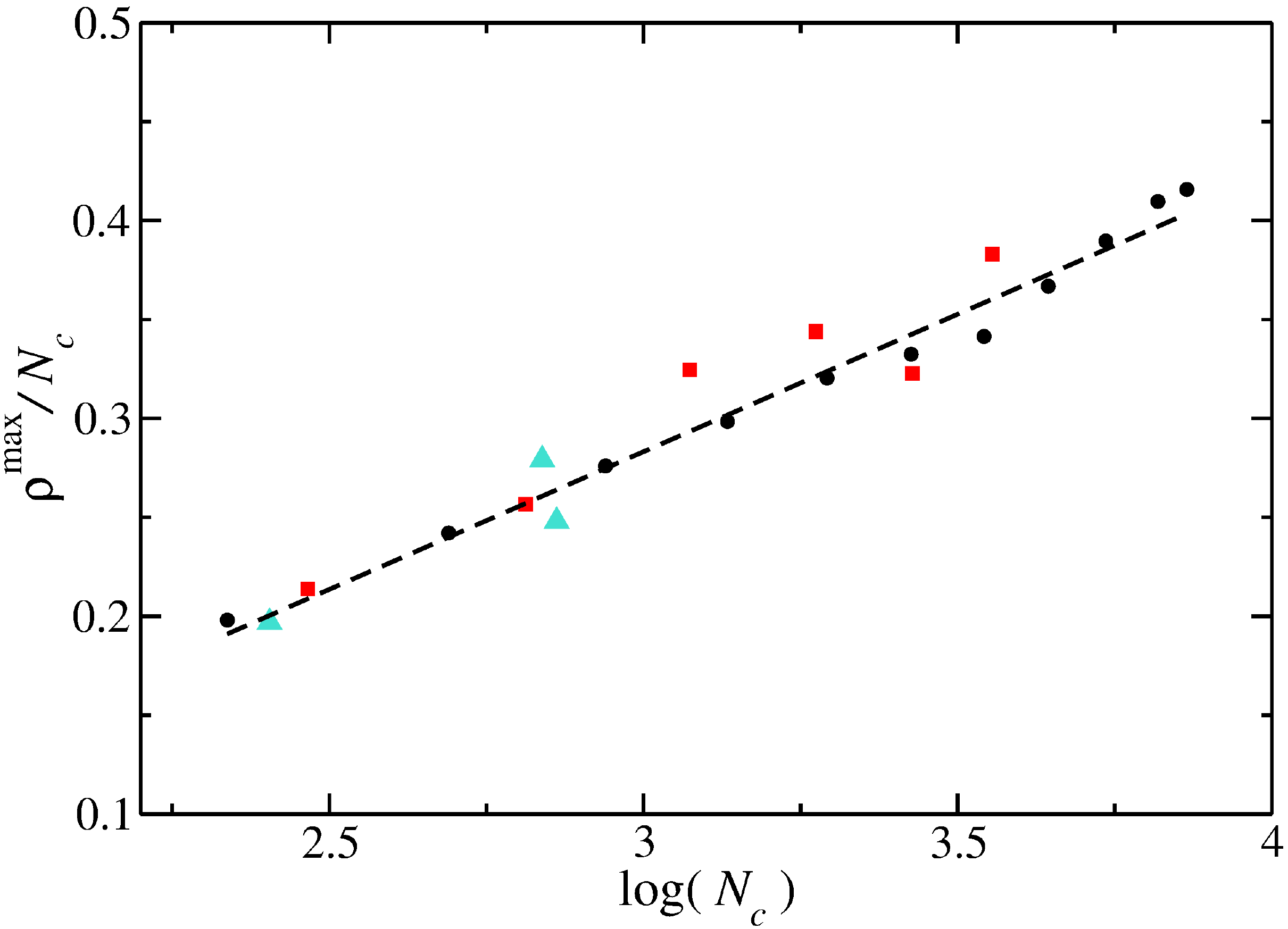}}
\caption{The height $\rho^{\rm max}$ of the van Hove peaks in the DOS of photonic crystals of varying size versus the logarithm of the number of hexagonal cells $N_c$. Black dots: Numerical results for photonic crystals with the shape of Africa. The dashed line shows a straight-line fit $a\log{(N_c)}+b$ through these data, where the resulting slope equaled $a=0.141$. Red squares: Numerical results for photonic crystals with the shape of a rectangle. Turquoise triangles: Experimental results.}
\label{fig23}
\end{figure}

The DOS diverges logarithmically at the van Hove singularities only for 2-dimensional, periodic structures of infinite extent. In the crystals used in the experiments, however, the sharp peaks in the DOS have a finite height $\rho^{\rm max}$. We determined $\rho^{\rm max}$ for the experimental DOS of the two microwave photonic crystals and a third, smaller one which contained only 267 cylinders, and also performed numerical studies for photonic crystals of various sizes with the shapes of a rectangle and of Africa. For a comparison of these results we rescaled the frequencies such that the distance between the van Hove singularities was 2 for all systems. 
The experimental and numerical studies revealed that the maxima of the DOS, $\rho^{\rm max}$, behave like
\begin{equation}
\rho^{\rm max}\simeq a N_c\left[\log(N_c)+b\right]\label{rhoexp}
\end{equation}
with $N_c$ denoting the number of unit cells, i.e., of hexagons formed by the voids in the photonic crystal. Here, $a$ and $b$ are fit parameters, where the former takes a similar value $a\sim 0.141-0.155\approx\frac{3}{2\pi^2}$ for all cases, i.e., it seems to be \emph{universal}.
\section{Conclusions\label{Conclusions}}
In this article, $1d$ (transverse out-of-plane) $z$-vibrations have been studied.
For the analysis of the experimental data, one needs to generalize the method to $2d$
(longitudinal in-plane) $xy$-vibrations. In the algebraic approach, this is done by
introducing the Lie algebra $g=u(3)$~\cite{Iachello1996,Iachello2009}. The combined algebra 
at each site is then $u(2)\oplus u(3)$. The application of the algebraic method to longitudinal
vibrations and to combinations of transverse and longitudinal vibrations will be
reported in a subsequent publication. Also, for those situations in which the
lattice is not composed of identical units $X-X-X-\dots$, but it has alternating units
$X-Y-X-Y-\dots$, one needs to compute the phonon dispersion relation both for the
optical and acoustic branches. This is easily done within the framework of the
algebraic theory discussed here.

Finally, we have presented a new method for calculating the energy
spectrum (EDR) and density of states (DOS) of vibrations of solids, both
harmonic and anharmonic, and applied it to the study of the EDRs and the DOSs of 1D linear chains,
and 2D square and hexagonal lattices. Our results have been compared with
data obtained in microwave photonic crystals for 2D hexagonal lattices.
These data show the expected occurrence of both van Hove singularities and
Dirac zeros. The method can be easily extended to other 2D-lattices, with
symmetry of the unit cell other than $D_{4h}$ and $D_{6h}$, and in fact to
any two-dimensional structure planar and non-planar, as for example
fullerene, C$_{60}$, with icosahedral symmetry $I_{h}$. The algebraic method
can also be used to calculate the response of a solid to infrared (IR) and
Raman (R) radiation. For 1D systems, this response was studied in \cite%
{iac-truini}. For 2D systems it remains to be done.

The method is also well suited to study the general Hubbard boson
Hamiltonian of Eq.~(\ref{eqn23}). Because of the formal equivalence between the boson
Hamiltonian, Eq.~(\ref{eqn57}), and the fermion Hamiltonian, Eq.~(\ref{eqn58}), it can also be
used to study the band structure in 1D, 2D square and hexagonal lattices in
the tight-binding model. The DOS of the fundamental vibration, $v=1$, of an
hexagonal lattice is identical to the DOS of electrons in the same lattice.

\section{Acknowledgements}

This work was supported by the DFG within the Collaborative Research Center 634. F.I. acknowledges
support from U.S.D.O.E. Grant DE-FG02-91ER40608.

\end{document}